\documentclass[ aps,pra,reprint,superscriptaddress,twocolumn,
showpacs,floatfix,footinbib]{revtex4-2}
\usepackage[utf8]{inputenc}
\usepackage{amsmath}
\usepackage{amsfonts}
\usepackage{dsfont} % added for mathds
\usepackage{bbm}
\usepackage{amssymb}
\usepackage{makeidx}
\usepackage{graphicx}
\usepackage{color}
\usepackage{mathptmx}
\usepackage{CJKutf8}
\usepackage{changes,enumitem,todonotes}
\usepackage{soul}
\usepackage[mathscr]{euscript}
\usepackage{tikz}
\usepackage[normalem]{ulem}
\usepackage[utf8]{inputenc}
\usepackage{amsmath}
\usepackage{amsfonts}
\usepackage{amssymb}
\usepackage{subcaption} % added for subcaption
\usepackage{hyperref}
\hypersetup{colorlinks=true,linkcolor=blue,urlcolor=blue,citecolor=blue} 
\usepackage{mathtools} % added for splitfrac
\newcommand{\RE}{\mathrm{Re}}

\newcommand{\orcid}[1]{\href{https://orcid.org/#1}
	{\includegraphics[width=7pt,height=7pt]{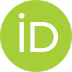}}}

\begin{document}
\title{Resonance-engineered single-photon routing in an asymmetric Jaynes-Cummings cavity-waveguide system}
\author{Adam S. Abdalla\,\orcid{0000-0002-5823-8789}}
\email[Author to whom correspondence should be addressed. Electronic mail: ]{adam.abdalla@physik.tu-darmstadt.de}
\affiliation{Institute of Applied Physics, Technical University of Darmstadt, Darmstadt 64289, Germany}%
\affiliation{Department of Physics, Faculty of Science, University of Kordofan, Elobeid 51111, Sudan}%

\begin{abstract}
We investigate single-photon scattering and routing in an asymmetric cavity-waveguide quantum electrodynamics system consisting of a Jaynes-Cummings cavity-emitter subsystem coupled to two one-dimensional waveguides. Using a real-space scattering approach, we derive an exact three-channel scattering matrix and analytical expressions for the channel-resolved scattering amplitudes. The scattering response is governed by the interplay of asymmetric cavity-waveguide coupling, atom-cavity detuning, coherent preparation of a single-photon superposition across multiple input channels, and intrinsic dissipation. Coupling asymmetry controls the relative weights of the cavity-mediated scattering pathways, whereas atom-cavity detuning modifies their spectral response. Coherent multichannel input states provide an additional phase-sensitive control of the output probabilities through interference. In particular, we identify an analytical condition, $J_2=2J_1$ and $\alpha_1=-\alpha_2$, under which the probability in the second output channel becomes independent of photon energy; for an equal-weight normalized input, this probability is 1/2. We further show how cavity and atomic dissipation reduce the probability remaining in the guided channels. These results provide a framework for resonance-engineered single-photon routing in asymmetric Jaynes-Cummings cavity-waveguide systems.      
\end{abstract}

\maketitle

%%%%%%%%%%%%%%%%%%%%%%%%%%%%%%%%%%%%%%%%%%%%%%%%%%%%%%%%%%
%%%%%%%%%%%%%%%%%%%%%%%%%%%%%%%%%%%%%%%%%%%%%%%%%%%%%%%%%%
%\section{Introduction}
\section{\label{sec:intro} Introduction}
%%%%%%%%%%%%%%%%%%%%%%%%%%%%%%%%%%%%%%%%%%%%%%%%%%%%%%%%%%
%%%%%%%%%%%%%%%%%%%%%%%%%%%%%%%%%%%%%%%%%%%%%%%%%%%%%%%%%%

Waveguide quantum electrodynamics (QED) provides a versatile platform for studying and controlling the interaction of propagating photons with localized quantum systems in effectively one-dimensional photonic environments \cite{Roy2017,Chang2018,Gu2017,Sheremet2023}. Strong light-matter coupling and engineered photon propagation in these systems enable controlled single-photon scattering, transmission, reflection, and routing, with applications in quantum communication, quantum networks, quantum information processing, quantum metrology, and integrated photonics \cite{Krenn2016communication,Couteau2023,Kuhn2002,Lu2014,Reiserer2015cavity,Monroe2002,Kiraz2004dot,Flamini2019photonic,Northup2014,Giovannetti2006,Dietric2016,Rodt2021}.

Cavity-waveguide QED provides a particularly useful architecture in which a localized cavity-mediated interaction is coupled to propagating photonic channels. A cavity containing a two-level emitter realizes a  Jaynes-Cummings subsystem that modifies the scattering of incident photons through resonant light-matter interaction. Such hybrid cavity-waveguide platforms have been investigated in photonic-crystal and nanophotonic structures, superconducting circuits, and atom- and quantum-dot-based systems \cite{Zhou2008,Jiang2019simulation,Zhou2008coherent,Tudela2024light,Wang2025giant,Santori2010nanophotonics,Abdelrahman2014,Katsumi2025,Wang2022simulating,Ozawa2019topological,Perczel2020}.
The coexistence of direct waveguide propagation and cavity-mediated scattering produces frequency-dependent interference and resonance phenomena, enabling control of transmission, reflection, switching, and photon routing \cite{Lalanne2018,Fleischhauer2005,Lei2022,Liu2017eit,Miroshnichenko2010fano,Limonov2017fano,Manasi2018,Vinu2023,Purdy2008,Carmele2013,Yan2015,Teklu2018,Mirhosseini2019cavity,Ahumada2019,Hoi2013,Li2024}. In multichannel configurations, the relative cavity-waveguide coupling strengths provide an additional degree of freedom for controlling the redistribution of a photon among the available output channels.

A central question in multichannel waveguide-QED scattering is how the properties of the scattering device and the coherent preparation of the incident photon jointly determine the output state. Previous studies have considered single-channel excitation, symmetric or otherwise fixed coupling configurations, and a variety of specific routing mechanisms \cite{Tudela2024light,Bermel2006,Li2024,Wu2024,Shen2009theory1,Shen2009theory2,Dong2019,Abdalla2020,Shen2005coherent,ShenFan2007,Duda2026,Li2015design,Lechner2023,Luo2023,Bello2019,Lan2024,Dong2013,Fratini2014,Abdalla2025,Bradford2013,Intaraprasonk2012,Bag2023,Pletyukhov2015,Astafiev2010}. A unified analytical treatment that incorporates both coupling asymmetry and coherent preparation of multiple input channels within the same Jaynes-Cummings scattering problem is useful for separating device-controlled scattering from input-state-controlled interference. In particular, when several input channels are coherently populated, the output probabilities depend not only on the intrinsic scattering amplitudes of the device but also on the relative amplitudes and phases of the incident single-photon state. 
\begin{figure}
\includegraphics[width=0.47\textwidth]{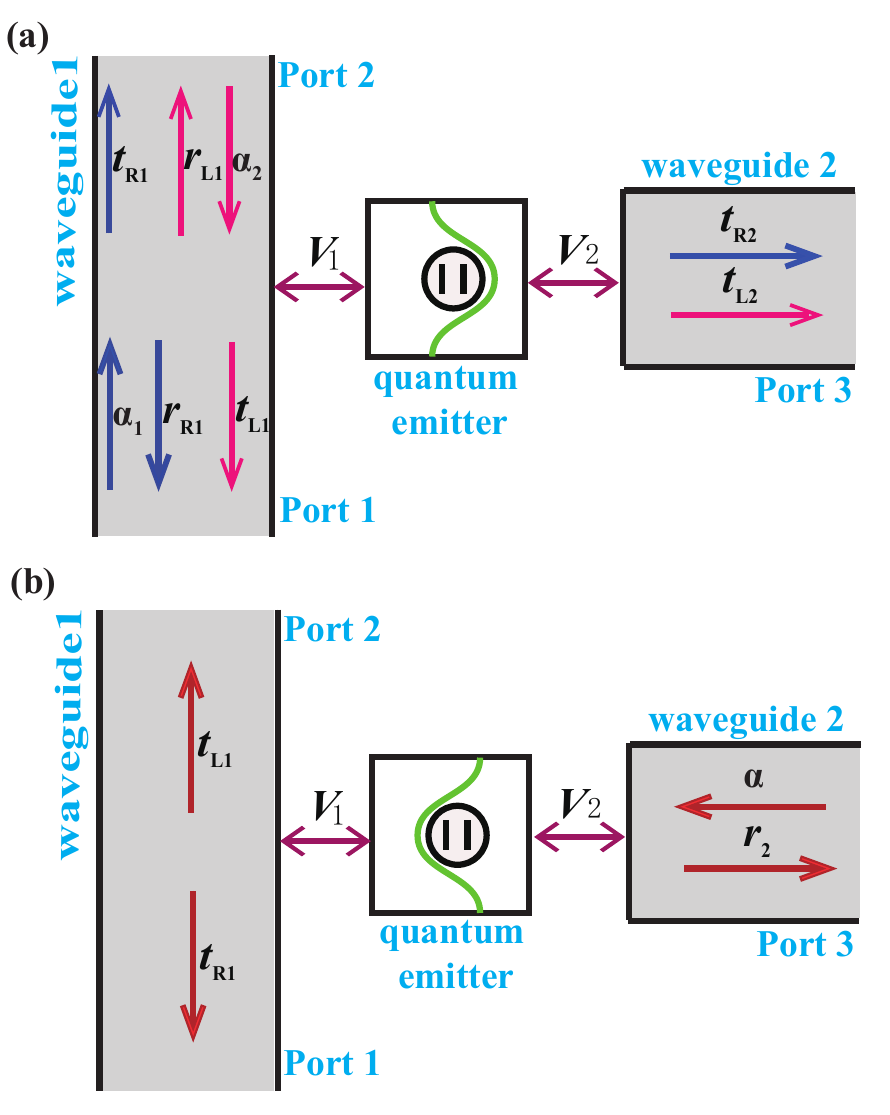}
\caption{(Color online) Schematic of the asymmetric Jaynes-Cummings cavity-waveguide system. The cavity containing a two-level emitter is side-coupled to waveguide 1 with strength of $V_{1}$ and end-coupled to waveguide 2 with strength of $V_{2}$. (a) Single-photon injection through either counterpropagating channel of waveguide 1 with input amplitudes $\alpha_1$ and $\alpha_2$. (b) Single-photon injection through port 3 of waveguide 2, with amplitude $\alpha$. The quantities $r$ and $t$ denote reflection and transmission amplitudes, respectively.}
\label{fig:fig0-structure}
\end{figure}

In this work, we develop a real-space scattering theory for single-photon transport in an asymmetric Jaynes-Cummings cavity-waveguide system consisting of two one-dimensional waveguides coupled to a common cavity-emitter subsystem. The cavity is side-coupled to a bidirectional waveguide and end-coupled to a semi-infinite waveguide, resulting in a three-channel scattering problem characterized by two independently tunable effective coupling strengths, $J_1$ and $J_2$. We derive the exact $3\times3$ scattering matrix and use it to analyze channel-resolved routing under single-channel and coherent multichannel excitation. We examine the effects of coupling asymmetry, atom-cavity detuning, coherent input-state preparation, and cavity and atomic dissipation. In particular, we identify an analytical condition, $J_2=2J_1$ and $\alpha_1=-\alpha_2$, under which the probability in the second output channel is independent of photon energy. We further show how coherent preparation of the input state provides an additional phase-sensitive control of the output probabilities through interference. These results establish a unified framework for analyzing and controlling single-photon routing in asymmetric Jaynes-Cummings cavity-waveguide systems. 

The remainder of the paper is organized as follows. Sec.~\ref{sec:model} ~introduces the cavity-waveguide model and develops the real-space single-photon scattering formalism. Sec.~\ref{sec:analytics} presents numerical results for asymmetric and symmetric coupling, coherent multichannel input states, dissipation, and atom-cavity detuning. Sec
.~\ref{sec:conc}~summarizes the main results.

%%%%%%%%%%%%%%%%%%%%%%%%%%%%%%%%%%%%%%%%%%%%%%%%%%%%%%%%%%
%%%%%%%%%%%%%%%%%%%%%%%%%%%%%%%%%%%%%%%%%%%%%%%%%%%%%%%%%%
\section{\label{sec:model}Model and theoretical framework}
%%%%%%%%%%%%%%%%%%%%%%%%%%%%%%%%%%%%%%%%%%%%%%%%%%%%%%%%%%
%%%%%%%%%%%%%%%%%%%%%%%%%%%%%%%%%%%%%%%%%%%%%%%%%%%%%%%%%%

\subsection{Physical model}

The asymmetric quantum-router architecture considered in this work is illustrated in Fig.~\ref{fig:fig0-structure}. The system consists of two one-dimensional waveguides coupled through a Jaynes-Cummings cavity-emitter subsystem, comprising a single-mode optical cavity containing a two-level quantum emitter. The cavity is side-coupled to waveguide 1 with coupling strength $V_{1}$ and end-coupled to waveguide 2 with coupling strength $V_{2}$. The two coupling strengths can be varied independently and therefore provide the structural parameters required to engineer asymmetric photon-scattering pathways.

We consider two complementary single-photon injection configurations. In Fig.~\ref{fig:fig0-structure}(a), the photon is injected into either of the two counterpropagating channels of waveguide 1, with input amplitudes $\alpha_1$ and $\alpha_2$. After interacting with the Jaynes-Cummings cavity-emitter subsystem, the photon can remain in waveguide 1 or be transferred to waveguide 2 through the cavity-mediated interaction. In Fig.~\ref{fig:fig0-structure}(b), the photon is injected through port 3 of waveguide 2 with amplitude $\alpha$. It can be reflected back into waveguide 2 or transmitted into either counterpropagating channel of waveguide 1. We neglect direct photon transfer between the two waveguides so that interwaveguide transport occurs exclusively through the Jaynes-Cummings cavity-emitter subsystem. This assumption isolates the cavity-mediated scattering process and allows the effects of the two cavity-waveguide coupling strengths to be studied independently.

The objective is to determine how the coupling asymmetry, resonance conditions, input-state preparation, and dissipation influence the redistribution of single-photon probability among the available output channels. The relative coupling strengths determine the weights of the cavity-mediated scattering pathways, while the cavity and atomic frequencies determine their spectral response. For coherent excitation of multiple input channels, the complex input amplitudes, including their relative phases, provide an additional control over the interference between scattering pathways. Cavity and atomic losses introduce additional dissipative channels and therefore reduce the probability remaining in the guided outputs.

We formulate this problem using a real-space single-photon scattering approach. The resulting scattering matrix provides a compact description of reflection, transmission, and interwaveguide routing. The derivation of the scattering amplitudes for the general asymmetric case is given in Appendix~\ref{app:deriv}, while the main text introduces the scattering matrix and uses it to formulate the channel-resolved routing probabilities.

%%%%%%%%%%%%%%%%%%%%%%%%%%%%%%%%%%%%%%%%%%%%%%%%%%%%%%%%%%
%%%%
\subsection{Hamiltonian and single-photon scattering equations}
%%%%
%%%%%%%%%%%%%%%%%%%%%%%%%%%%%%%%%%%%%%%%%%%%%%%%%%%%%%%%%%

%The total Hamiltonian is written as \cite{Shen2009theory1,Shen2009theory2,Abdalla2020,Abdalla2025}
%\begin{align}
%H=H_{W1}+H_{W2}+H_{JC}+H_{I1}+H_{I2},
%\label{eq:totalHamil}
%\end{align}
The total Hamiltonian is written as $H=H_{W1}+H_{W2}+H_{end}+H_{JC}+H_{I1}+H_{I2}$,
where $H_{W1}$ and $H_{W2}$ describe the free propagation of photons in the two waveguides, $H_{end}$ describes the endpoint interaction in the waveguide 2, $H_{JC}$ describes the cavity-atom subsystem, and $H_{I1}$ and $H_{I2}$ describe the corresponding cavity-waveguide interactions. The full Hamiltonian in real space is given by \cite{Shen2009theory1,Shen2005coherent,Abdalla2020,Abdalla2025}:

\begin{widetext}
\begin{subequations}
\label{eq:componHamil}
\begin{align}
&H_{W_1} =\int dy\hat\psi_{L1}^\dag (y)\left(\omega_0+iv_g \frac {\partial}{\partial y}\right) \hat\psi_{L1}(y)+\int dy {\hat\psi_{R1}}^\dag (y) \left(\omega_0-iv_g \frac {\partial}{\partial y}\right) \hat\psi_{R1}(y)\\
&H_{W_2} =\int_{0}^\infty dx\hat\psi_{L2}^\dag (x)\left(\omega_0+iv_g \frac {\partial}{\partial x}\right) \hat\psi_{L2}(x)+\int_{0}^\infty dx {\hat\psi_{R2}}^\dag (x) \left(\omega_0-iv_g \frac {\partial}{\partial x}\right) \hat\psi_{R2}(x)\\
&H_{end} =\eta\int_{0}^\infty dx\delta(x){\hat\psi_{R2}}^\dag (x){\hat\psi_{L2}} (x)+\eta^*\int_{0}^\infty dx\delta(x){\hat\psi_{L2}}^\dag (x){\hat\psi_{R2}} (x)\\
&H_{I_1} =\int dy V_1\delta(y)\hat a^+\left[\hat\psi_{R1}(y)+\hat\psi_{L1}(y)\right]+\int dy V_1\delta(y)\hat a\left[\hat\psi_{R1}^\dag (y)+\hat\psi_{L1}^\dag (y)\right]\\
&H_{I_2} =\int_{0}^\infty dx V_2\delta(x)\hat a^+\left[\hat\psi_{R2}(x)+\hat\psi_{L2}(x)\right]+\int_{0}^\infty dx V_2\delta(x)\hat a\left[\hat\psi_{R2}^\dag (x)+\hat\psi_{L2}^\dag (x)\right]\\
&H_{JC} =\tilde{\omega}_c\hat a^+\hat a +\tilde{\omega}_a\sigma_+\sigma_-+\Omega(\hat a^+\hat\sigma_-+\hat\sigma_+\hat a)
%&H_{JC} =\omega_c\hat a^+\hat a +\Omega(\hat a^+\hat \sigma^- +\hat a\hat \sigma^+)+ \omega_e \hat a_e^\dag \hat a_e + \omega_g \hat a_g^\dag \hat a_g
\end{align}
\end{subequations}
\end{widetext}
The waveguide modes are treated within a linear-dispersion approximation about the reference frequency $\omega_0$, while the cavity-waveguide interactions are represented by localized $\delta$-function couplings, i.e., $\varepsilon(k)=\omega, \omega=\omega_0+v_gk$, where $v_g$ and $k$ are the group velocity and wave vector, respectively. The right- and left-propagating modes are described by the corresponding real-space field operators, and the cavity-waveguide interactions are modeled by localized $\delta$-function couplings of strengths $V_1$ and $V_2$. We introduce an auxiliary endpoint interaction $\eta$ to implement the boundary condition at the termination of the semi-infinite waveguide [see Appendix~\ref{app:deriv}]. This effective description imposes the termination boundary condition while retaining the cavity-mediated coupling to the guided mode.

The Jaynes-Cummings cavity-emitter subsystem is characterized by cavity frequency $\omega_c$, atomic transition frequency $\omega_a$, where the atom-cavity detuning is $\Delta_{ac}=\omega_a-\omega_c$, and cavity-atom coupling $\Omega$. Dissipation is incorporated through  non-Hermitian frequencies $\tilde{\omega}_c=\omega_c-i\gamma_c$ and $\tilde{\omega}_a=\omega_a-i\gamma_a$, where $\gamma_c$ and $\gamma_a$  are defined here as amplitude-decay parameters. Because the non-Hermitian frequencies are written as $\omega_j-i\gamma_j$, the corresponding population decay rates are $2\gamma_j$.   Cavity and atomic dissipation reduce the probability remaining in the guided channels, which are quantified by $P_{guided}=P_1+P_2+P_3$ and $P_{loss}=1-P_{guided}$.

For a single excitation with energy $\varepsilon$, we use the real-space ansatz 
%~\cite{b54,b55}.
\begin{align}
|\Phi\rangle &= \int dy\phi_{R1}(y){\hat\psi_{R1}}^\dag(y)|\varnothing\rangle
+ \int dy\phi_{L1}(y){\hat\psi_{L1}}^\dag(y)|\varnothing\rangle\nonumber\\
&+\int_{0}^\infty dx\phi_{R2}(x){\hat\psi_{R2}}^\dag(x)|\varnothing\rangle
+ \int_{0}^\infty dx\phi_{L2}(x){\hat\psi_{L2}}^\dag(x)|\varnothing\rangle\nonumber\\
&+A\hat\sigma^+|\varnothing\rangle
+C\hat a^+|\varnothing\rangle
\label{eq:ansatz}
\end{align}
%%%%%%%%%%%%%%%%%%%%%%%%%%%%%%%%%%%%%%%%%%%%%%%%%
\begin{figure}
\includegraphics[width=0.47\textwidth]{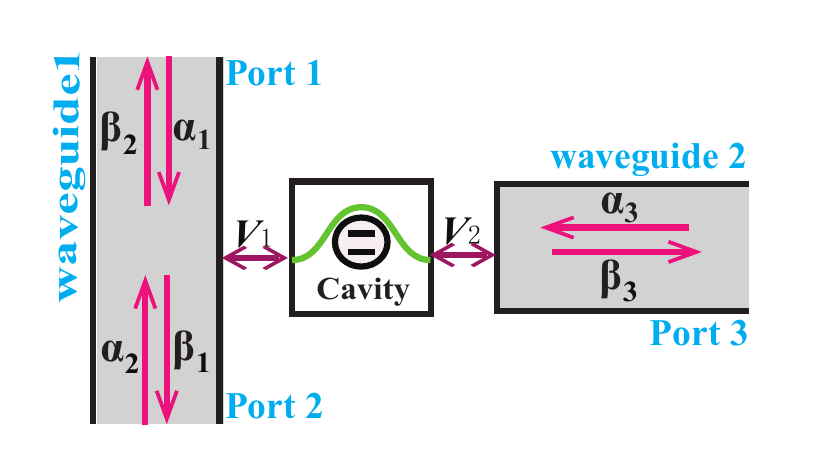}
\caption{(Color online) Continuous asymmetric Jaynes-Cummings cavity-waveguide system representation of general input $\alpha_i$ and output $\beta_i$ amplitudes.}
\label{fig:fig1-structure}
\end{figure}
%%%%%%%%%%%%%%%%%%%%%%%%%%%%%%%%%%%%%%%%%%%%%%%%%

In the single-excitation sector, we employ the real-space scattering ansatz of Eqs.~\eqref{eq:ansatz}. Substitution into stationary Schr\"{o}dinger equation $H|\Phi\rangle=\varepsilon|\Phi\rangle$ yields the coupled equations for the four propagating photon amplitudes and the atomic and cavity excitation amplitudes. The boundary conditions associated with these equations determine the outgoing amplitudes for arbitrary combinations of incoming fields.
%\begin{widetext}
\begin{subequations}
\label{eq:dynamicseq}
\begin{align}
&-iv_g \frac {\partial}{\partial y}\phi_{R1}(y)+CV_1\delta(y)=(\varepsilon-\omega_0)\phi_{R1}(y)\\
&iv_g \frac {\partial}{\partial y}\phi_{L1}(y)+CV_1\delta(y)=(\varepsilon-\omega_0)\phi_{L1}(y)\\
&-iv_g \frac {\partial}{\partial x}\phi_{R2}(x)+CV_2\delta(x)+\eta\phi_{L2}(x)\delta(x)
=(\varepsilon-\omega_0)\phi_{R2}(x)\\
&iv_g \frac {\partial}{\partial x}\phi_{L2}(x)+CV_2\delta(x)+\eta^*\phi_{R2}(x)\delta(x)
=(\varepsilon-\omega_0)\phi_{L2}(x)\\
&V_1\left[\phi_{R1}(0)+\phi_{L1}(0)\right]+V_2\left[\phi_{R2}(0)+\phi_{L2}(0)\right]+\Omega A
=\tilde{\omega}_cC\\
&\Omega C=\tilde{\omega}_aA
\end{align}
\end{subequations}
%\end{widetext}  

%%%%%%%%%%%%%%%%%%%%%%%%%%%%%%%%%%%%%%%%%%%%%%%%%
\begin{figure*}
		\begin{center}
			{\includegraphics[width=0.90\textwidth]{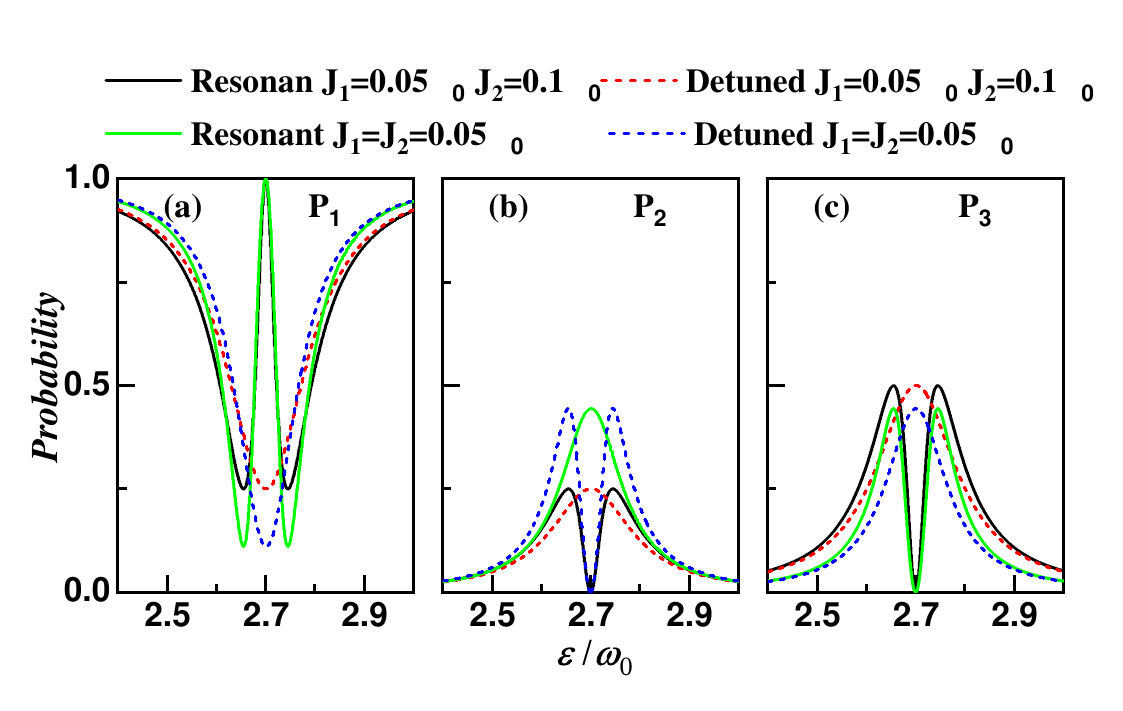}}
			\caption{(Color online) Output probabilities $P_1$, $P_2$, and $P_3$ as functions of the normalized photon energy $\varepsilon/\omega_0$ for a single-photon incident through port 1, corresponding to $|\alpha_1|^2=1$ and $\alpha_2=\alpha_3=0$. The resonant and detuned configurations are compared for asymmetric couplings $J_1=0.05\omega_0$, $J_2=0.1\omega_0$  and symmetric couplings $J_1=J_2=0.05\omega_0$ in the lossless limit.}
			\label{fig:Fig1-100}
		\end{center}
\end{figure*}
%%%%%%%%%%%%%%%%%%%%%%%%%%%%%%%%%%%%%%%%%%%%%%%%%

Rather than solving the equations separately for each injection configuration, we introduce the input and output amplitude vectors as illustrated in Fig.~\ref{fig:fig1-structure}.
\begin{align}
\bm \alpha_{in}=
\left(\begin{array}{c}
 \alpha_1 \\
 \alpha_2 \\
 \alpha_3
\end{array}\right),~~~~~~
\bm \beta_{out}
=\left(\begin{array}{c}
 \beta_1 \\
 \beta_2 \\
 \beta_3
\end{array}\right)
\label{eq:vectors}
\end{align}
the resulting expression gives the scattering relation
\begin{align}
\bm \beta_{out}=\bm {S}(\varepsilon)\bm \alpha_{in}
\label{eq:scmatrix}
\end{align}
with a $3\times3$ scattering matrix $\bm {S}(\varepsilon)$,
\begin{align}
\bm {S}(\varepsilon)
=\frac{1}{D(\varepsilon)}\left(\begin{array}{ccc}
  A(\varepsilon)+\frac{iJ_2}{2} &  -iJ_1 &  -i\sqrt{J_1J_2} \\
  -iJ_1 & A(\varepsilon)+\frac{iJ_2}{2} &  -i\sqrt{J_1J_2} \\
  -i\sqrt{J_1J_2} &  -i\sqrt{J_1J_2} &  A(\varepsilon)+iJ_1-\frac{iJ_2}{2}
\end{array}\right)
\label{eq:scatteringmatrix}
\end{align}
%\begin{align}
%\bm {S}(\varepsilon)
%=\left(\begin{array}{ccc}
  %S_{11} &  S_{12} &  S_{13} \\
  %S_{21} &  S_{22} &  S_{23} \\
  %S_{31} &  S_{32} &  S_{33}
%\end{array}\right)
%\label{eq:scatteringmatrix}
%\end{align}
where the matrix elements $\bm {S}_{ij}(\varepsilon)$ are derived explicitly in Appendix~\ref{app:deriv}. Here, $J_1=V_1^2/v_g$ and $J_2=V_2^2/v_g$ are the effective cavity-waveguide coupling strengths. Furthermore, we can define pole parameter $D(\varepsilon)$ governing the asymmetric effective couplings, atom-cavity resonance detuning, and dissipative losses as
\begin{align}
D(\varepsilon)=A(\varepsilon)+iJ_1+\frac{iJ_2}{2}
\label{eq:poleparameter}
\end{align}
%\begin{align}
%D(\varepsilon)=\varepsilon-\tilde{\omega}_c+iJ_1+\frac{iJ_2}{2}-\frac{\Omega^2}{\varepsilon-\tilde{\omega}_a}
%\label{eq:poleparameter}
%\end{align}
with 
\begin{align}
A(\varepsilon)=\varepsilon-\tilde{\omega}_c-\frac{\Omega^2}{\varepsilon-\tilde{\omega}_a}
\label{eq:parameter}
\end{align}
the same expression applies to both the lossless and dissipative calculations, with $\tilde{\omega}_{c,a}\rightarrow\omega_{c,a}$ in the lossless limit $\gamma_c=\gamma_a=0$. The poles of the scattering matrix are given by the zeros of $D(\varepsilon)$, 
\begin{align}
\left(\varepsilon-\tilde{\omega}_c+iJ_1+\frac{iJ_2}{2}\right)(\varepsilon-\tilde{\omega}_a)-\Omega^2=0
\label{eq:zerothpoints}
\end{align}
these poles correspond to the complex dressed resonances of the coupled cavity-emitter system, including the waveguide-induced broadening and the intrinsic dissipative losses. In the isolated, lossless Jaynes-Cummings limit, $J_1=J_2=\gamma_a=\gamma_c=0$, the corresponding dressed-state frequencies reduce to  
\begin{align}
\omega_\pm=\frac{\omega_c+\omega_a}{2}\pm\frac{1}{2}\sqrt{(\omega_a-\omega_c^2)^2+4\Omega^2}
\label{eq:eigenfrequencies}
\end{align}
Using the scattering matrix in Eq. $\eqref{eq:scatteringmatrix}$, the probability in output channel $i$ for an input state normalized according to $\sum_j|\alpha_j|^2=1$ is  
\begin{align}
P_i=|\beta_i|^2=|\sum_j\bm {S}_{ij}\alpha_j|^2.
\label{eq:genproba}
\end{align}
For a coherently prepared single-photon superposition across multiple input channels, Eq. $\eqref{eq:genproba}$ can be expanded as
\begin{align}
P_i=\sum_j|\bm {S}_{ij}|^2|\alpha_j|^2+2\RE{\left[\sum_{j<k}\bm {S}_{ij}\bm {S}_{ik}^\ast\alpha_j\alpha_k^\ast\right]}
\label{eq:simproba}
\end{align}
the second term represents the interference between scattering pathways associated with different input channels. Thus, both the magnitudes and relative phases of the incident amplitudes can influence the output routing probabilities. We can write the cross term explicitly as
\begin{align}
P_i^{int}=2\RE{\left[\sum_{j<k}\bm {S}_{ij}\bm {S}_{ik}^\ast\alpha_j\alpha_k^\ast\right]}.
\label{eq:crossterm}
\end{align}
The first term of Eq. $\eqref{eq:simproba}$ represents the sum of the individual channel contributions, while the second term contains the interference contributions arising from a coherently prepared single-photon superposition across multiple input channels. To achieve the proposed model we consider four scattering configurations to solve the dynamics set Eqs. $\eqref{eq:dynamicseq}$ based on forms of the wave functions, namely.

%%%%%%%%%%%%%%%%%%%%%%%%%%%%%%%%%%%%%%%%%%%%%%%%%
\begin{figure}
\includegraphics[width=0.47\textwidth]{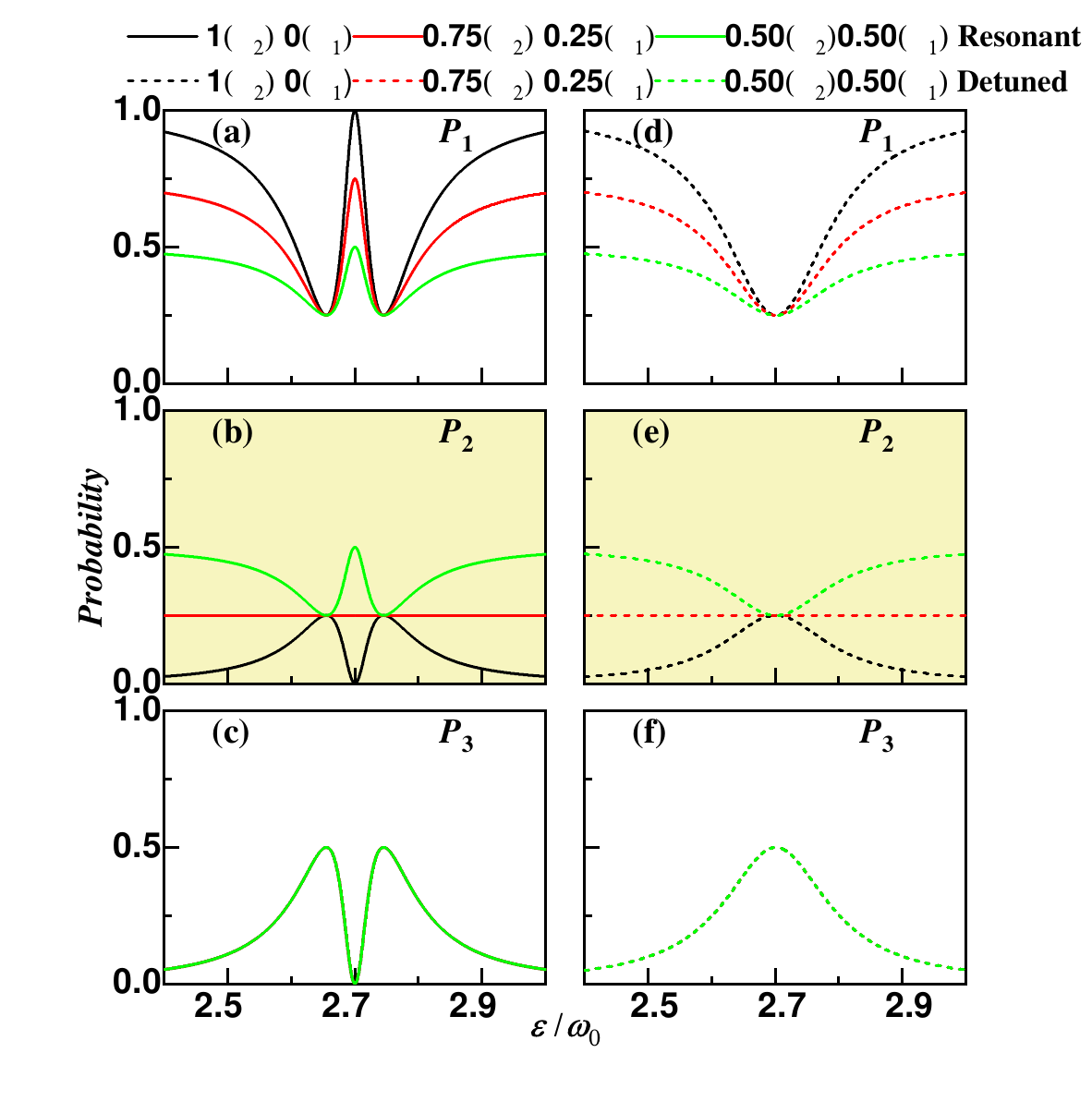}
\caption{(Color online) Output probabilities $P_1$, $P_2$, and $P_3$ as functions of the normalized photon energy $\varepsilon/\omega_0$ for a single photon prepared in a coherent superposition of input channels 1 and 2 with amplitudes $\alpha_1$ and $\alpha_2$.  The plotted values for relative input amplitudes ($a_1, a_2$)= (1, 0), (0.75, 0.25), (1, -1), and (0.5, 0.5) for asymmetric couplings $J_1=0.005\omega_0$ and $J_2=0.01\omega_0$ in the lossless limit.}
\label{fig:fig3-J1notJ2-fromport12}
\end{figure}
%%%%%%%%%%%%%%%%%%%%%%%%%%%%%%%%%%%%%%%%%%%%%%%%%

%--------------------------------------------------
\subsubsection{Input through port 1}
%--------------------------------------------------

For photon input-state $\bm \alpha_{in}=(\alpha_1, 0, 0)^\top$, Eq. $\eqref{eq:genproba}$ ~gives
\begin{align}
P_1=|S_{11}\alpha_1|^2, ~ P_2=|S_{21}\alpha_1|^2, ~P_3=| S_{31}\alpha_1|^2.
\label{eq:gproba1}
\end{align}
using the amplitude definitions shown in Fig.~\ref{fig:fig0-structure}(a),
\begin{subequations}
\begin{align}
\phi_{R1}(y)&= r_{R1}e^{iky}\theta(y)\\
\phi_{L1}(y)&=\ \alpha_{1}e^{-iky}\theta(y)+t_{R1}e^{-iky}\theta(-y)\\
\phi_{R2}(x)&= t_{R2}e^{ikx}\theta(x)\\
\phi_{L2}(x)&=0
\end{align}
\label{eq:wf1}
\end{subequations}
for a unit-amplitude input Eq. $\eqref{eq:gproba1}$ we obtain
\begin{align}
P_1=|t_{R1}|^2|\alpha_1|^2, ~ P_2=|r_{R1}|^2|\alpha_1|^2, ~P_3=|t_{R2}|^2|\alpha_1|^2.
\label{eq:proba1}
\end{align}

%--------------------------------------------------
\subsubsection{Input through port 2}
%--------------------------------------------------

For a unit-amplitude input $\bm \alpha_{in}=(0, \alpha_2, 0)^\top$ as defined in Fig.~\ref{fig:fig0-structure}(a),
\begin{subequations}
\label{eq:wf2}
\begin{align}
\phi_{R1}(y)&= r_{R1}e^{iky}\theta(y)\\
\phi_{L1}(y)&=\ \alpha_{2}e^{-iky}\theta(y)+t_{R1}e^{-iky}\theta(-y)\\
\phi_{R2}(x)&= t_{R2}e^{ikx}\theta(x)\\
\phi_{L2}(x)&=0
\end{align}
\end{subequations}
Eq. $\eqref{eq:genproba}$ gives
\begin{align}
P_1=|S_{12}\alpha_2|^2, ~ P_2=|S_{22}\alpha_2|^2, ~P_3=|S_{32}\alpha_2|^2.
\label{eq:gproba2}
\end{align}
or, equivalently,
\begin{align}
P_1=|t_{L1}|^2|\alpha_2|^2, ~ P_2=|r_{L1}|^2|\alpha_2|^2, ~P_3=|t_{L2}|^2|\alpha_2|^2.
\label{eq:proba2}
\end{align}
%%%%%%%%%%%%%%%%%%%%%%%%%%%%%%%%%%%%%%%%%%%%%%%%%
\begin{figure*}
		%\begin{center}
        \centering
			{\includegraphics[width=1.0\textwidth]{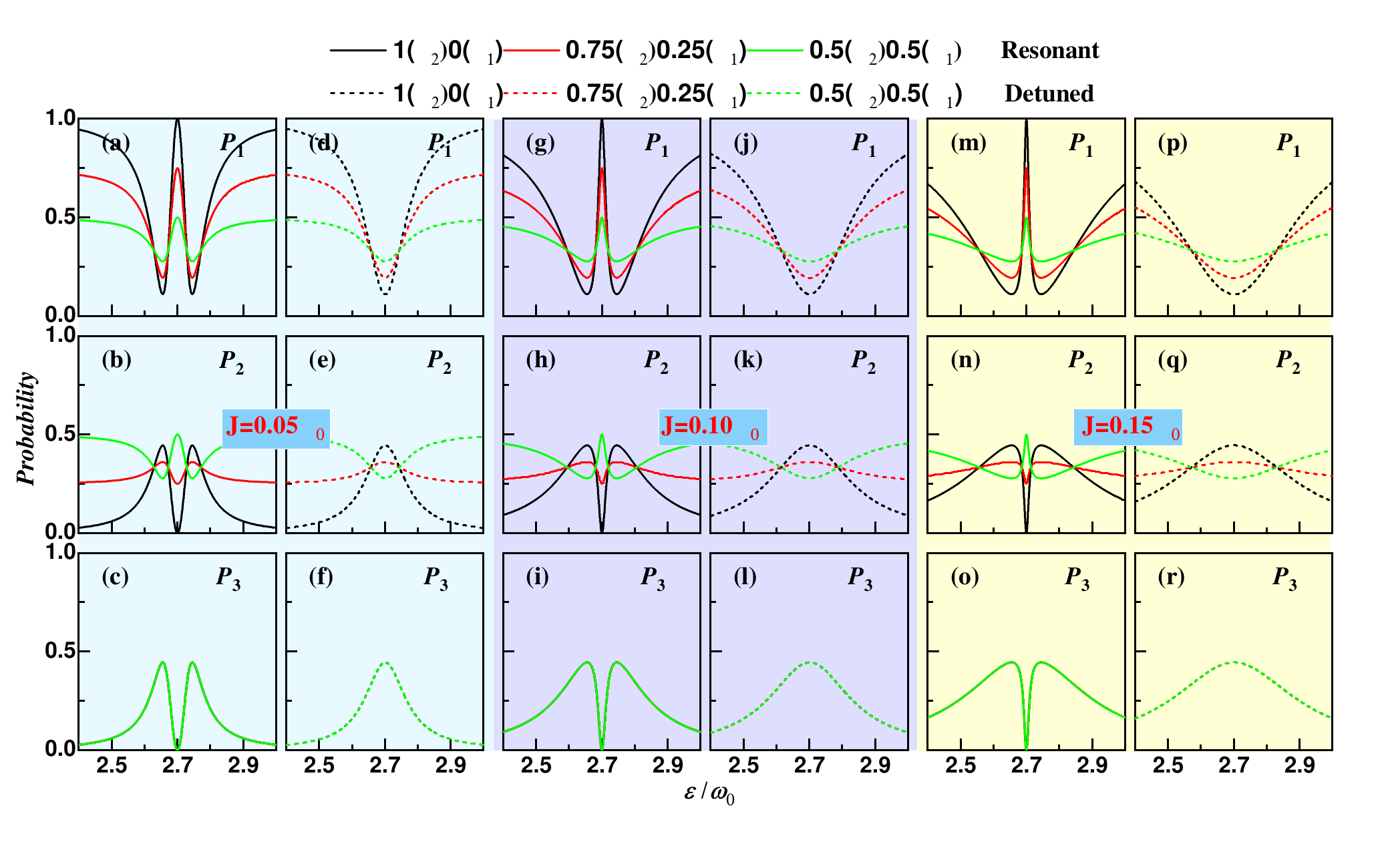}}
			\caption{(Color online) Output probabilities $P_1$, $P_2$, and $P_3$ as functions of the normalized photon energy $\varepsilon/\omega_0$ for a single photon prepared in a coherent superposition of input channels 1 and 2 with amplitudes $\alpha_1$ and $\alpha_2$.  The plotted values for relative input amplitudes ($a_1, a_2$)= (1, 0), (0.75, 0.25), (1, -1), and (0.5, 0.5) for symmetric couplings $J=0.05\omega_0$ in (a, b, c, d, e, f), $J=0.10\omega_0$ in (g, h, i, j, k, l), and $J=0.15\omega_0$ in (m, n, o, p, q, r) in the lossless limit.}
			\label{fig:fig4-J-fromport12}
		%\end{center}
	\end{figure*}
%%%%%%%%%%%%%%%%%%%%%%%%%%%%%%%%%%%%%%%%%%%%%%%%%

%--------------------------------------------------
\subsubsection{Coherent excitation of ports 1 and 2}
%\textbf{(iii) Coherent excitation of ports 1 and 2}\\
%--------------------------------------------------

For a simultaneous coherently prepared single-photon superposition across multiple input channels $\bm \alpha_{in}=(\alpha_1, \alpha_2, 0)^\top$, the three output probabilities are
\begin{subequations}
\label{eq:gproba3}
\begin{align}
P_1=|S_{11}\alpha_1+S_{12}\alpha_2|^2,\\
P_2=|S_{21}\alpha_1+S_{22}\alpha_2|^2,\\
P_3=|S_{31}\alpha_1+S_{32}\alpha_2|^2.
\end{align}
\end{subequations}
in terms of the scattering amplitudes,
\begin{subequations}
\label{eq:proba3}
\begin{align}
P_1=|t_{R1}\alpha_1+t_{L1}\alpha_2|^2,\\
P_2=|r_{R1}\alpha_1+r_{L1}\alpha_2|^2,\\
P_3=|t_{R2}\alpha_1+t_{L2}\alpha_2|^2.
\end{align}
\end{subequations}
These expressions make the interference mechanism completely explicit as Eq. $\eqref{eq:simproba}$.

%%%%%%%%%%%%%%%%%%%%%%%%%%%%%%%%%%%%%%%%%%%%%%%%%
\begin{figure*}
		%\begin{center}
        \centering
			{\includegraphics[width=0.80\textwidth]{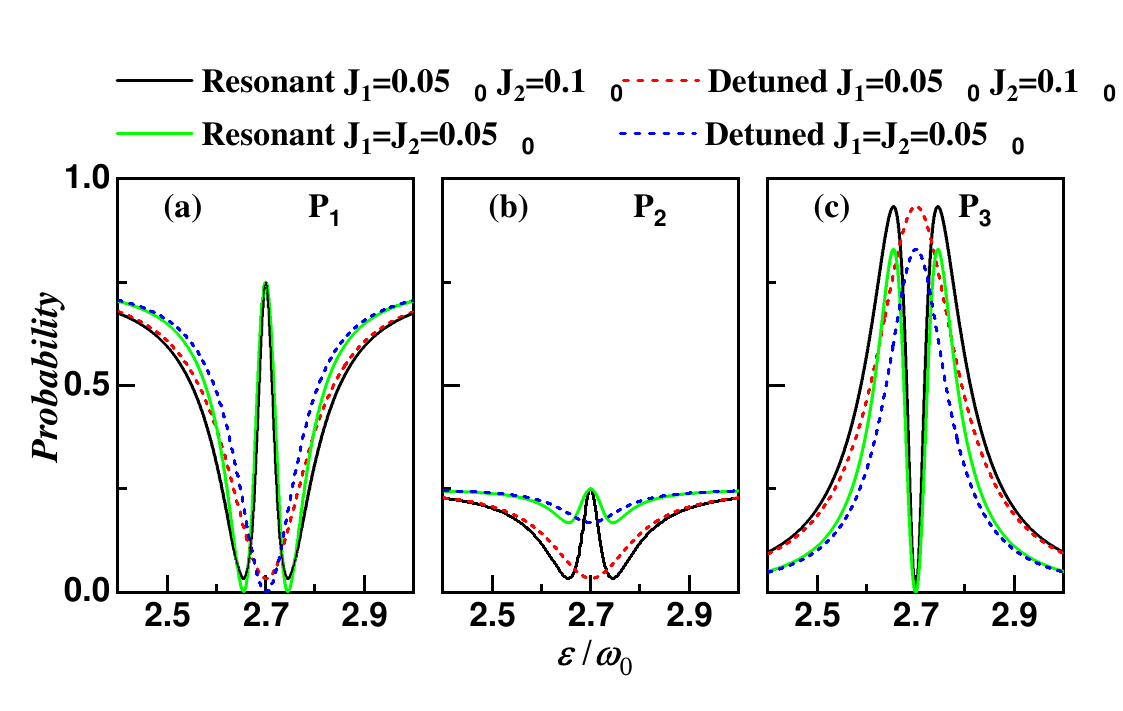}}
			\caption{(Color online) Output probabilities $P_1$, $P_2$, and $P_3$ as functions of the normalized photon energy $\varepsilon/\omega_0$ for a single-photon prepared in a coherent superposition of input channels 1 and 2, with normalized input amplitudes $\alpha_1=\sqrt{3}/2$, and $\alpha_2=1/2$ and $\alpha_3=0$ with zero relative phase. The resonant and detuned configurations are compared for asymmetric couplings $J_1=0.05\omega_0$, $J_2=0.1\omega_0$  and symmetric couplings $J_1=J_2=0.05\omega_0$ in the lossless limit.}
			\label{fig:Fig2-75250}
		%\end{center}
\end{figure*}
%%%%%%%%%%%%%%%%%%%%%%%%%%%%%%%%%%%%%%%%%%%%%%%%%

%--------------------------------------------------
\subsubsection{Input through port 3}
%--------------------------------------------------

For an input through port 3, with $\bm \alpha_{in}=(0, 0, \alpha_3)^\top$ as defined in Fig.~\ref{fig:fig0-structure}(b),
\begin{subequations}
\label{eq:wf4}
\begin{align}
\phi_{R1}(y)&= t_{R1}e^{iky}\theta(y)\\
\phi_{L1}(y)&= t_{L1}e^{-iky}\theta(-y)\\
\phi_{R2}(x)&= r_{2}e^{ikx}\theta(x)\\
\phi_{L2}(x)&=\alpha e^{-ikx}\theta(x)
\end{align}
\end{subequations}
Eq. $\eqref{eq:genproba}$ gives
\begin{align}
P_1=|S_{13}\alpha_3|^2, ~ P_2=|S_{23}\alpha_3|^2, ~P_3=|S_{33}\alpha_3|^2.
\label{eq:gproba4}
\end{align}
and equivalently,
\begin{align}
P_1=|t_{R1}|^2|\alpha_3|^2, ~ P_2=|t_{L1}|^2|\alpha_3|^2, ~P_3=|r_{2}|^2|\alpha_3|^2.
\label{eq:proba4}
\end{align}

%%%%%%%%%%%%%%%%%%%%%%%%%%%%%%%%%%%%%%%%%%%%%%%%%
\begin{figure*}
		%\begin{center}
        \centering
			{\includegraphics[width=0.80\textwidth]{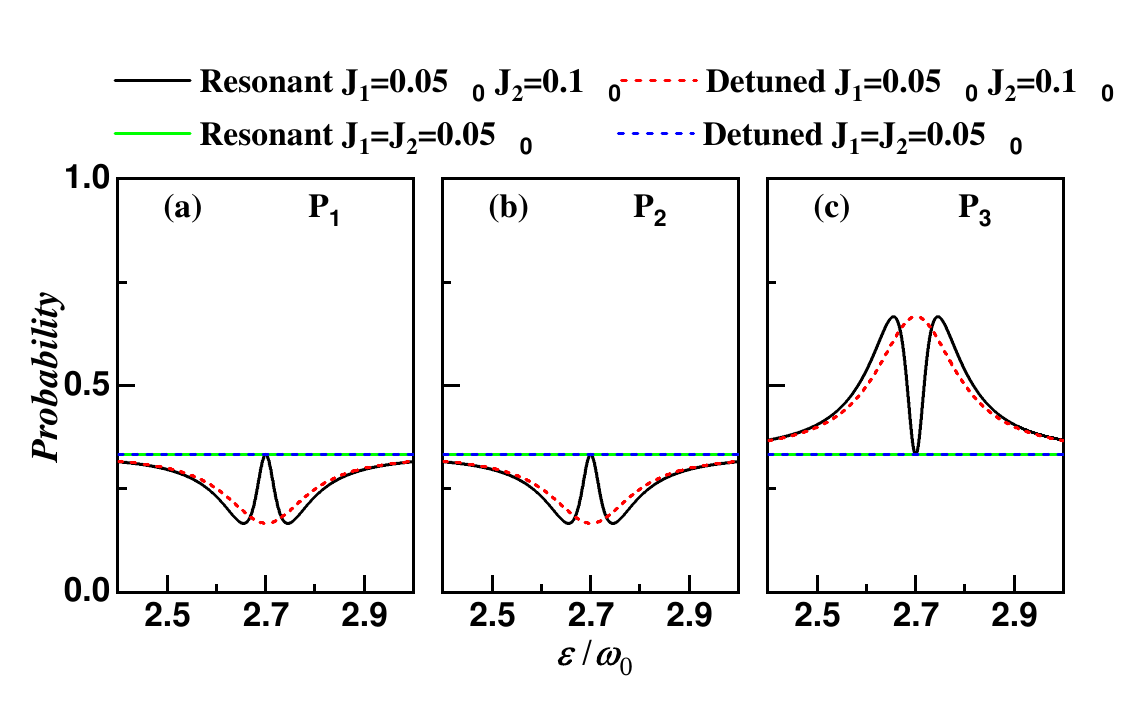}}
			\caption{(Color online) Output probabilities $P_1$, $P_2$, and $P_3$ as functions of the normalized photon energy $\varepsilon/\omega_0$ for a single-photon prepared in an equal-weight coherent superposition of the  three input channels, $\alpha_1=\alpha_2=\alpha_3=1/\sqrt{3}$. The resonant and detuned configurations are compared for asymmetric couplings $J_1=0.05\omega_0$, $J_2=0.1\omega_0$  and symmetric couplings $J_1=J_2=0.05\omega_0$ in the lossless limit.}
			\label{fig:Fig4-333}
		%\end{center}
\end{figure*}
%%%%%%%%%%%%%%%%%%%%%%%%%%%%%%%%%%%%%%%%%%%%%%%%%
    
%--------------------------------------------------
\subsubsection{Coherent excitation of all three input channels}
%--------------------------------------------------

For a single photon prepared in a coherent superposition of the three input channels, $\bm \alpha_{in}=(\alpha_1, \alpha_2, \alpha_3)^\top$, as defined in Fig.~\ref{fig:fig1-structure}, the general single-photon scattering ansatz
\begin{subequations}
\begin{align}
\phi_{R1}(y)&= \alpha_{1}e^{iky}\theta(-y)+\beta_1e^{iky}\theta(y)\\
\phi_{L1}(y)&= \alpha_{2}e^{-iky}\theta(y)+\beta_{2}e^{-iky}\theta(-y)\\
\phi_{R2}(x)&= \beta_{3}e^{ikx}\theta(x)\\
\phi_{L2}(x)&= \alpha_3e^{-ikx}\theta(x)
\end{align}
\label{eq:spansatz}
\end{subequations}
For a coherent superposition of the three input channels, the incident single-photon state is
\begin{align}
|\Phi_{in}\rangle =\alpha_1|1, 0, 0\rangle+\alpha_2|0, 1, 0\rangle+\alpha_3|0, 0, 1\rangle
\label{eq:coherentstate}
\end{align}
To specify this state, one must define the input probabilities $|\alpha_j|^ 2$ and relative phase $\phi_j-\phi_k$.  The corresponding output probabilities are given by Eq. $\eqref{eq:genproba}$.
\begin{subequations}
\begin{align}
P_1=|S_{11}\alpha_1+S_{12}\alpha_2+S_{13}\alpha_3|^2,\\
P_2=|S_{21}\alpha_1+S_{22}\alpha_2+S_{23}\alpha_3|^2,\\
P_3=|S_{31}\alpha_1+S_{32}\alpha_2+S_{33}\alpha_3|^2.
\end{align}
\label{eq:superpositionproba123}
\end{subequations}

%%%%%%%%%%%%%%%%%%%%%%%%%%%%%%%%%%%%%%%%%%%%%%%%%%%%%%%%%%
%%%%%%%%%%%%%%%%%%%%%%%%%%%%%%%%%%%%%%%%%%%%%%%%%%%%%%%%%%
\section{\label{sec:analytics} Numerical results and analysis}
%\section{Numerical verification and analysis}
%%%%%%%%%%%%%%%%%%%%%%%%%%%%%%%%%%%%%%%%%%%%%%%%%%%%%%%%%%
%%%%%%%%%%%%%%%%%%%%%%%%%%%%%%%%%%%%%%%%%%%%%%%%%%%%%%%%%%

The scattering amplitudes derived in Sec.~\ref{sec:model} and Appendix~\ref{app:deriv} determine the channel-resolved output probabilities for arbitrary single-photon input states. In this section, we use the scattering matrix to examine four aspects of the transport response: (i) structural asymmetry of the cavity-waveguide couplings, (ii) coherent preparation of the incident photon across multiple input channels, (iii) cavity and atomic dissipation, and (iv) atom-cavity detuning. Unless otherwise stated, the calculations use $\omega_0=2\pi\times10^{10}Hz$, $\Omega=0.045\omega_0$, $\omega_c=2.7\omega_0$, with $\lambda_0=3~cm$,  $k_0=2\pi/\lambda_0$, $v_g=0.6 \omega_0/k_0$. The resonant and detuned configurations correspond to $\omega_a=2.7\omega_0$ and $0.7\omega_0$, respectively \cite{Shen2009theory1,Shen2005coherent,Abdalla2020,Abdalla2025}.
Unless otherwise stated, the calculations are performed in the lossless limit when the effects of coupling asymmetry, coherent input preparation, or detuning are considered separately.

For a set of unnormalized relative input amplitudes $a_j$, we define the normalized single-photon amplitudes by $\alpha_j=a_j/\sqrt{\sum_k|a_k|^2}$, $\sum_j|\alpha_j|^2=1$. The output probabilities are then calculated from $P_i=|\sum_j\bm {S}_{ij}\alpha_j|^2$. For multichannel coherent input state, the relative phases of the amplitudes are part of the physical specification of the incident state and are stated explicitly whenever they affect the interference response.
%%%%%%%%%%%%%%%%%%%%%%%%%%%%%%%%%%%%%%%%%%%%%%%%%
\begin{figure*}
		%\begin{center}
        \centering
			{\includegraphics[width=0.96\textwidth]{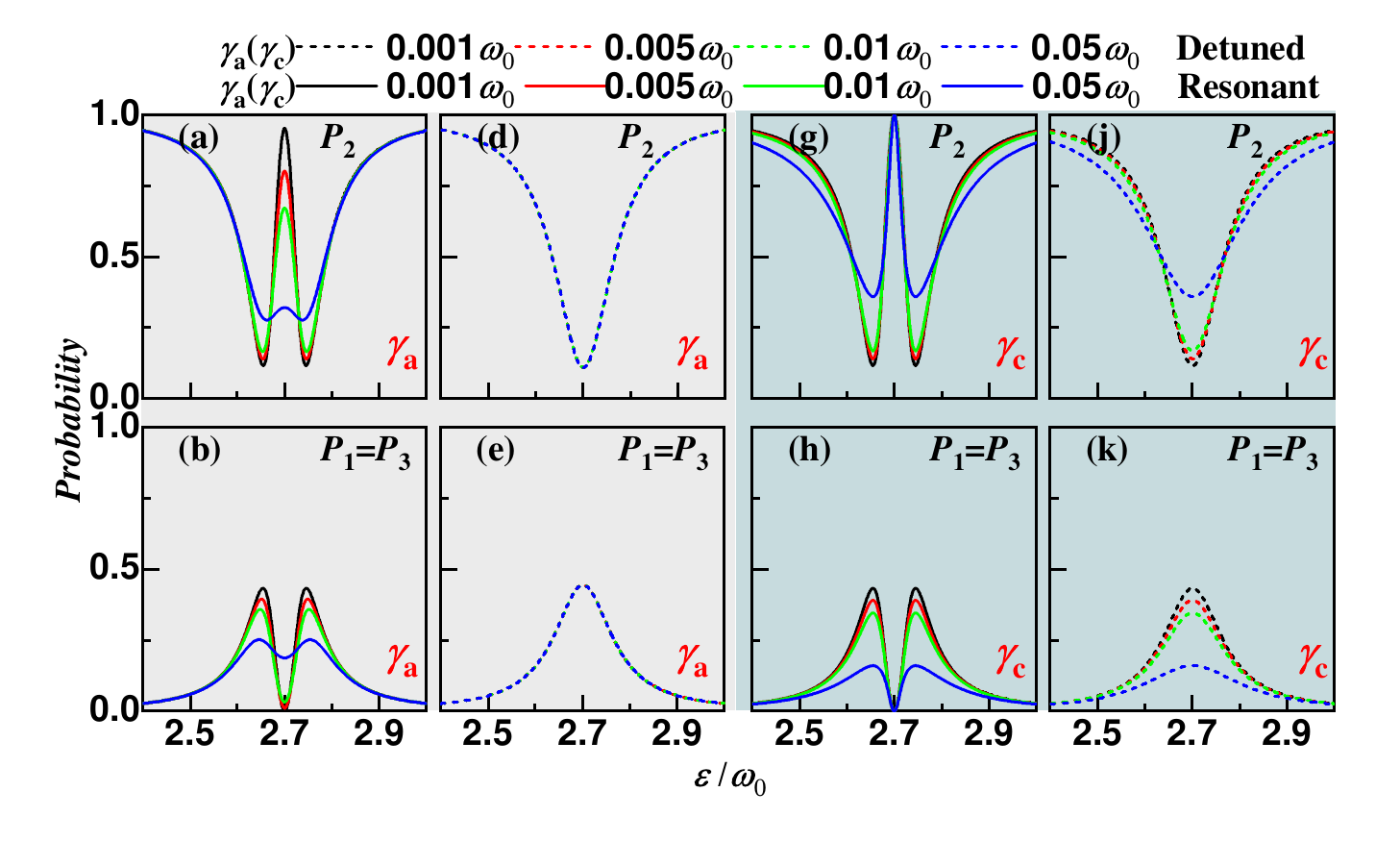}}
			\caption{(Color online) Output probabilities $P_1$, $P_2$, and $P_3$ as functions of the normalized photon energy $\varepsilon/\omega_0$ for a single photon incident through port 2, corresponding to $|\alpha_2|^ 2=1$ and $\alpha_1=0$. The plotted values for atomic decay parameter $\gamma_a = 0.001\omega_0, 0.005\omega_0, 0.01\omega_0, 0.05\omega_0$ and $\gamma_c=0$ in (a, b, c, d) and for cavity decay parameter $\gamma_c = 0.001\omega_0, 0.005\omega_0, 0.01\omega_0, 0.05\omega_0$ and $\gamma_a=0$ in (e, f, g, h) for symmetric couplings $J_1=J_2=0.005\omega_0$.}
			\label{fig:fig5-J-fromport1201-loss}
		%\end{center}
\end{figure*}
%%%%%%%%%%%%%%%%%%%%%%%%%%%%%%%%%%%%%%%%%%%%%%%%%

%%%%%%%%%%%%%%%%%%%%%%%%%%%%%%%%%%%%%%%%%%%%%%%%%%%%%%%%%%  
\subsection{Effect of asymmetric cavity-waveguide coupling}
%%%%%%%%%%%%%%%%%%%%%%%%%%%%%%%%%%%%%%%%%%%%%%%%%%%%%%%%%%

We first examine the role of structural coupling asymmetry in the single-photon scattering response. We consider a photon incident through port 1 and compare an asymmetric configuration, $J_1=0.005\omega_0$ and $J_2=0.01\omega_0$, with symmetric configurations $J_1=J_2=J$, where $J=0.05\omega_0$, $0.10\omega_0$, and $0.15\omega_0$. All calculations in this subsection are performed in the lossless limit to isolate the effect of the cavity-waveguide coupling strengths. The corresponding spectra are shown in Fig.~\ref{fig:Fig1-100}.
For symmetric coupling, the two waveguides interact with the cavity with equal effective coupling strength. In the asymmetric configuration, the relative coupling strengths modify the weights of the cavity-mediated scattering pathways, thereby changing the magnitudes and phases of the output amplitudes and the resulting spectral redistribution among the output channels. Coupling asymmetry therefore provides a structural control parameter that is distinct from atom-cavity detuning. Whereas $J_1$ and $J_2$ determine the relative coupling of the cavity to the waveguides, $\Delta_{ac}$ modifies the spectral response of the coupled cavity-emitter subsystem. These two parameters therefore provide complementary mechanisms for controlling channel-resolved single-photon scattering.

%%%%%%%%%%%%%%%%%%%%%%%%%%%%%%%%%%%%%%%%%%%%%%%%%%%%%%%%%%
\begin{figure*}
		\begin{center}
        %\centering
			{\includegraphics[width=1.1\textwidth]{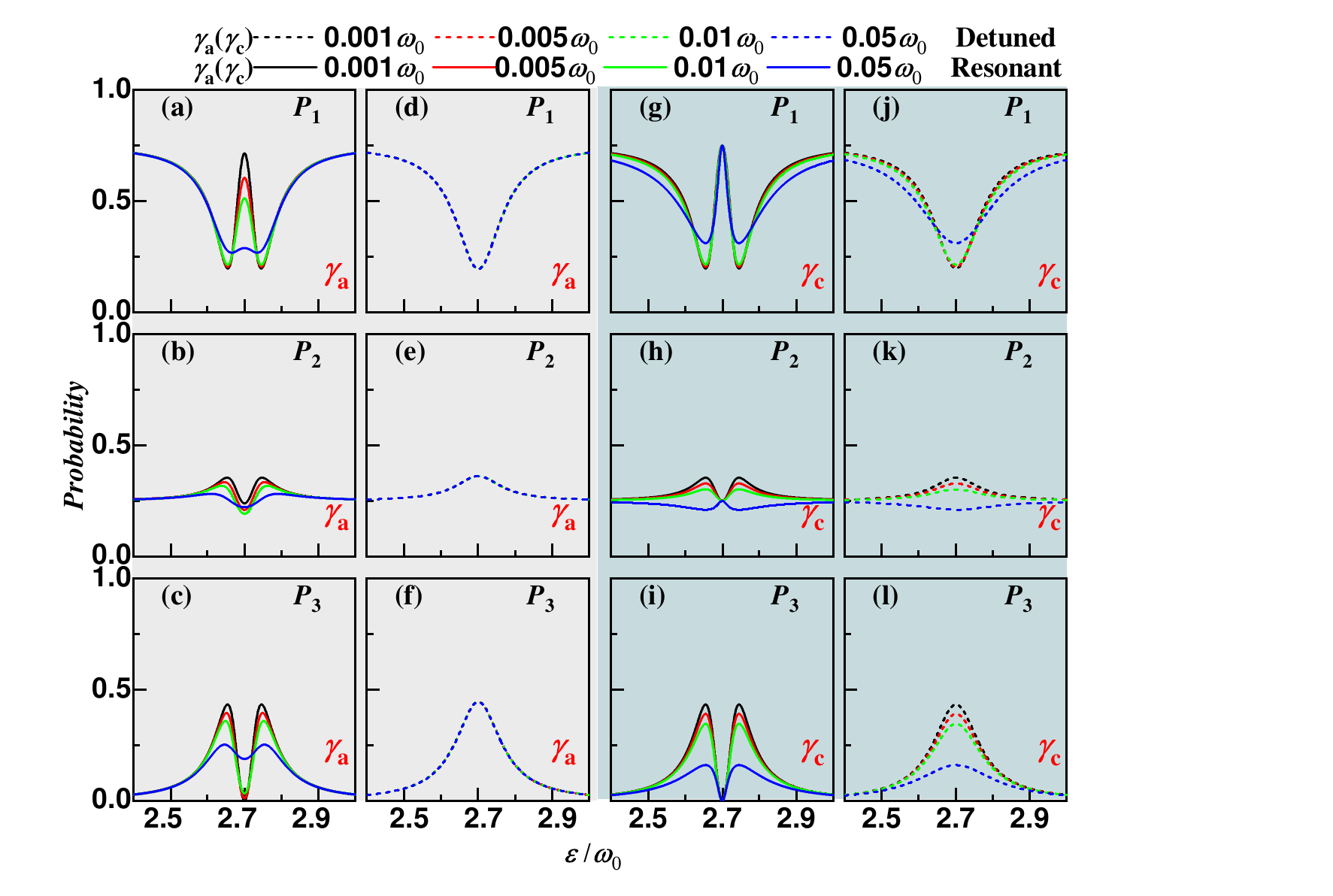}}
			\caption{(Color online) Output-channel probabilities $P_1$, $P_2$, and $P_3$ as functions of the normalized photon energy $\varepsilon/\omega_0$ for a single-photon prepared in a coherent superposition of input channels 1 and 2 with normalized input amplitudes $\alpha_1=1/2$, and $\alpha_2=\sqrt{3}/2$ and $\alpha_3=0$ with zero relative phase. The results compare different atomic decay parameter $\gamma_a$ (a, b, c, d, e, f) and different cavity decay parameter $\gamma_c$ (g, h, i, j, k, l) for symmetric couplings $J_1=J_2=0.005\omega_0$.}
			\label{fig:fig6-J-fromp12-2575-loss}
		\end{center}
\end{figure*}
%%%%%%%%%%%%%%%%%%%%%%%%%%%%%%%%%%%%%%%%%%%%%%%%%%%%%%%%%%

%%%%%%%%%%%%%%%%%%%%%%%%%%%%%%%%%%%%%%%%%%%%%%%%%%%%%%%%%%
\subsection{Influence of a coherently prepared single-photon superposition across multiple input channels}
%%%%%%%%%%%%%%%%%%%%%%%%%%%%%%%%%%%%%%%%%%%%%%%%%%%%%%%%%%

For coherent excitation of ports 1 and 2, the output amplitude in port 2 is $\beta_2=S_{21}\alpha_1+S_{22}\alpha_2$. Using $S_{21} =-iJ_1/D(\varepsilon)$ and $S_{22} =(A(\varepsilon)+iJ_2/2)/D(\varepsilon)$, with $A(\varepsilon)=\varepsilon-\tilde{\omega}_c-\Omega^2/(\varepsilon-\tilde{\omega}_a)$, we find that the condition $J_2=2J_1$,  $\alpha_1=-\alpha_2$ and corresponds to a relative phase $\phi_1-\phi_2=\pi$ gives $\beta_2=-\alpha_1$. Consequently, $P_2=|\alpha_1|^2$, which is independent of the photon energy and of the atom-cavity detuning. For the normalized equal-weight input $\alpha_1=1/\sqrt{2}$,
$\alpha_2=-1/\sqrt{2}$, one obtains $P_2=1/2$.
This energy-independent output probability results from destructive interference between the two coherently prepared input pathways. Importantly, it arises from the preparation of the incident single-photon state rather than from a modification of the intrinsic scattering matrix. It therefore provides an analytical example of input-state control of single-photon routing.
 
Figures~\ref{fig:fig3-J1notJ2-fromport12}-\ref{fig:Fig2-75250} illustrate this effect for a single photon prepared in a coherent superposition of input channels 1 and 2, while Fig.~\ref{fig:Fig4-333} shows the scattering response for a single-photon prepared in an equal-weight coherent superposition of the three input channels. For the asymmetric configuration  $J_1=0.005\omega_0$ and $J_2=0.01\omega_0$, Fig.~\ref{fig:fig3-J1notJ2-fromport12} compares with symmetric cases $J=0.05\omega_0$, $0.10\omega_0$ and $0.15\omega_0$ in Fig.~\ref{fig:fig4-J-fromport12} for several relative input-amplitude configurations. The output spectra change substantially as the relative contribution of the two input channels is varied. This behavior follows directly from the cross term in Eq.~$\eqref{eq:crossterm}$, which can either enhance or suppress a given output channel depending on the relative phase and magnitude of the two incident amplitudes.

This behavior is a direct consequence of the interference between the two input pathways rather than a modification of the intrinsic scattering matrix. It therefore illustrates the distinction between device-controlled scattering and input-state-controlled interference. The condition $J_2=2J_1$ together with  $\alpha_1=-\alpha_2$ provides an analytically exact example of energy-independent routing. The spectra in Figs.~\ref{fig:fig3-J1notJ2-fromport12}-\ref{fig:Fig4-333} illustrate the more general dependence of the output probabilities on the relative amplitudes and phases of the input channels. Changing the common coupling strength in the symmetric configuration also modifies the spectral response to a given input-state preparation. The results demonstrate that the scattering properties of the device and the preparation of the incident state constitute separate, but complementary, control parameters. For the numerical calculations, we distinguish between relative input amplitudes and normalized single-photon amplitudes. Given a set of relative amplitudes $a_j$, the corresponding normalized input state is $\alpha_j=a_j/\sqrt{\sum_k|a_k|^2}$ and output probabilities are evaluated using normalized input states satisfying  $\sum_j|\alpha_j|^2=1$. For the three-channel coherent input considered in Fig.~\ref{fig:Fig4-333}, we choose equal real amplitudes, $\alpha_1=\alpha_2=\alpha_3=1/\sqrt{3}$, corresponding to equal input probabilities and zero relative phase. The numerical results in Fig.~\ref{fig:Fig4-333} are consistent with the analytical scattering matrix expression in Eq. $\eqref{eq:coherentstate}$ with the magnitudes $|\alpha_j|^2=1/3$ and relative phases $\alpha_j=e^ {i\phi_j}/\sqrt{3}$ and the same zero relative phases, followed in Fig.~\ref{fig:Fig2-75250} with magnitudes $|\alpha_1|^2=3/4$, $|\alpha_2|^2=1/4$. 

%%%%%%%%%%%%%%%%%%%%%%%%%%%%%%%%%%%%%%%%%%%%%%%%%
\begin{figure}
\includegraphics[width=0.49\textwidth]{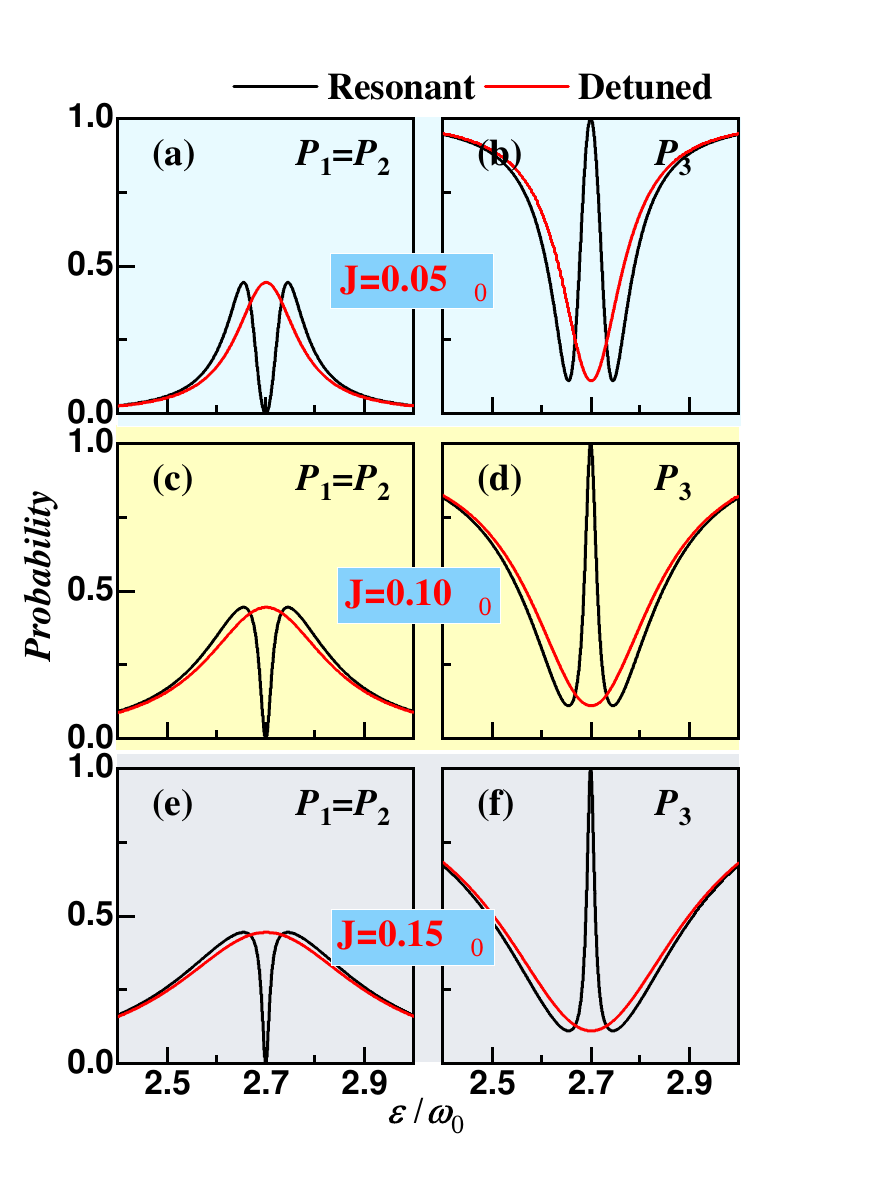}
\caption{(Color online) Output probabilities $P_1$, $P_2$, and $P_3$ as functions of the normalized photon energy $\varepsilon/\omega_0$ for a single photon incident through port 3, corresponding to $|\alpha_3|^ 2=1$. The resonant and detuned configurations are compared for symmetric couplings $J=0.05\omega_0$ (a, b), $J=0.10\omega_0$ (c, d), and $J=0.15\omega_0$ (e, f) in the lossless limit.}
\label{fig:fig9-J1=J2-fromport3}
\end{figure}
%%%%%%%%%%%%%%%%%%%%%%%%%%%%%%%%%%%%%%%%%%%%%%%%%

%%%%%%%%%%%%%%%%%%%%%%%%%%%%%%%%%%%%%%%%%%%%%%%%%%%%%%%%%%
\subsection{Dissipative effects}
%%%%%%%%%%%%%%%%%%%%%%%%%%%%%%%%%%%%%%%%%%%%%%%%%%%%%%%%%%

We next examine the effects of cavity and atomic dissipation on the guided scattering probabilities. Figures~\ref{fig:fig5-J-fromport1201-loss} and~\ref{fig:fig6-J-fromp12-2575-loss} show the output spectra for different values of atomic decay parameter $\gamma_a$ and cavity decay parameter $\gamma_c$. The total probability remaining in the three guided output channels is $P_{guided}=P_1+P_2+P_3$, while the probability lost from the guided channels is defined as $P_{loss}=1-P_{guided}$. In the lossless limit, $P_{guided}=1$. For nonzero $\gamma_a$ or $\gamma_c$, $P_{guided}<1$, reflecting the phenomenological dissipative channels represented by the non-Hermitian cavity and atomic frequencies.

For the detuned configuration considered here, the atomic transition is substantially displaced from the cavity resonance. Consequently, near the cavity resonance the atomic component contributes less strongly to the cavity-mediated response when $|\omega_a-\omega_c|\gg\Omega$. In the parameter regime investigated, cavity dissipation therefore produces a stronger modification of the guided-channel probabilities than atomic dissipation. This comparison is parameter dependent and should not be interpreted as a general hierarchy between the two loss mechanisms.

%%%%%%%%%%%%%%%%%%%%%%%%%%%%%%%%%%%%%%%%%%%%%%%%%%%%%%%%%%
\subsection{Single-photon routing under resonant and detuned conditions}
%%%%%%%%%%%%%%%%%%%%%%%%%%%%%%%%%%%%%%%%%%%%%%%%%%%%%%%%%%

We finally examine how atom-cavity detuning modifies the single-photon scattering response. In the resonant configuration, $\Delta_{ac}=0$ the cavity and atomic transition are resonant and form hybridized Jaynes-Cummings dressed states. The corresponding scattering poles produce characteristic spectral features associated with the coherent cavity-emitter coupling. 
When the cavity and atomic transition are detuned, $\Delta_{ac}\neq0$, the positions and linewidths of the complex dressed poles change, thereby modifying the frequency-dependent scattering amplitudes and channel-resolved output probabilities.
Detuning therefore provides a control parameter distinct from coupling asymmetry. Whereas $J_1$ and $J_2$ determine the relative coupling of the cavity to the waveguides, $\Delta_{ac}$ modifies the spectral response of the coupled cavity-emitter subsystem. Their combined variation provides independent control over the redistribution of single-photon probability among the available output channels. 

The coexistence of enhanced and suppressed output probabilities results from interference among the scattering pathways. Constructive interference enhances selected output channels, whereas destructive interference suppresses them.
Figure~\ref{fig:fig9-J1=J2-fromport3} extends this comparison to injection through port 3 for symmetric couplings. The corresponding spectra are described by the same scattering matrix: changing the input channel changes the relevant scattering amplitudes, while the cavity-emitter parameters continue to determine the underlying frequency-dependent response.
Taken together, the results demonstrate that structural coupling asymmetry, input-state preparation, dissipation, and atom-cavity detuning act on different aspects of the scattering response. Their combined use therefore provides a multivariable framework for controlling the redistribution of single-photon probability among the available output channels.

%%%%%%%%%%%%%%%%%%%%%%%%%%%%%%%%%%%%%%%%%%%%%%%%%%%%%%%%%%
%%%%%%%%%%%%%%%%%%%%%%%%%%%%%%%%%%%%%%%%%%%%%%%%%%%%%%%%%%
\section{\label{sec:conc} Conclusion}
%\section{conclusion}
%%%%%%%%%%%%%%%%%%%%%%%%%%%%%%%%%%%%%%%%%%%%%%%%%%%%%%%%%%
%%%%%%%%%%%%%%%%%%%%%%%%%%%%%%%%%%%%%%%%%%%%%%%%%%%%%%%%%%

We have developed a real-space scattering theory for single-photon transport in an asymmetric Jaynes-Cummings cavity-waveguide QED system. The cavity-emitter subsystem is coupled to a bidirectional waveguide and a terminated waveguide, resulting in three scattering channels characterized by two independently tunable effective coupling strengths, $J_1$ and $J_2$. Starting from the single-excitation Schr\"{o}dinger equation, we derived an exact $3\times3$ scattering matrix that provides a unified description of reflection, transmission, and interwaveguide routing.
The scattering-matrix approach separates the intrinsic response of the device from the preparation of the incident photon. The coupling strengths $J_1$ and $J_2$ determine the relative weights of the cavity-mediated scattering pathways, whereas atom-cavity detuning modifies their spectral response through the dressed cavity-emitter resonances. For coherent excitation of multiple input channels, the relative amplitudes and phases of the incident single-photon state provide an additional control of the output probabilities through interference. 
A particularly simple analytical result occurs for $J_2=2J_1$ and $\alpha_1=-\alpha_2$, for which the probability in the second output channel becomes independent of photon energy. For an equal-weight normalized input, this probability is $P_2=1/2$. This result demonstrates that coherent preparation of the incident photon can provide routing control beyond the spectral response determined by the intrinsic scattering properties of the device.
We further investigated coherent excitation of all three input channels and showed that the same scattering-matrix framework describes the  resulting multichannel interference. Finally, cavity and atomic dissipation reduce the probability remaining in the guided channels, with the relative influence depending on the atom-cavity detuning and the parameter regime. The framework developed here can be extended to more general multichannel waveguide networks, additional dissipative channels, direct interwaveguide coupling, and dynamically tunable system parameters. 

%%%%%%%%%%%%%%%%%%%%%%%%%%%%%%%%%%%%%%%%%%%%%%%%%%%%%%%%%%
%\begin{acknowledgements}
\section*{Acknowledgement}
The authors are grateful to Professor Dr. Reinhold Walser for fruitful discussions.
This research was supported by the HessenFonds Fellowship for Researchers at the Risk (1.8.2025-31.7.2026) and DLR German Aerospace Center with funds provided by the Federal Ministry for Economic Affairs and Climate Action (BMWK) due to an enactment for the German Bundestag under Grant No. 50WM2250E (QUANTUS+) at the Technical University of Darmstadt, Germany.
%\end{acknowledgements}
%%%%%%%%%%%%%%%%%%%%%%%%%%%%%%%%%%%%%%%%%%%%%%%%%%%%%%%%%%

%%%%%%%%%%%%%%%%%%%%%%%%%%%%%%%%%%%%%%%%%%%%%%%%%%%%%%%%%%
%\begin{author declarations}
\section*{author declarations}
The authors declare no conflicts of interest.
%\end{author declarations}
%%%%%%%%%%%%%%%%%%%%%%%%%%%%%%%%%%%%%%%%%%%%%%%%%%%%%%%%%%

%%%%%%%%%%%%%%%%%%%%%%%%%%%%%%%%%%%%%%%%%%%%%%%%%%%%%%%%%%
%\begin{Data availability}
\section*{Data availability}
The data that support the findings of this study are available from the corresponding author upon reasonable request.
%\end{Data availability}
%%%%%%%%%%%%%%%%%%%%%%%%%%%%%%%%%%%%%%%%%%%%%%%%%%%%%%%%%%

%%%%%%%%%%%%%%%%%%%%%%%%%%%%%%%%%%%%%%%%%%%%%%%%%%%%%%%%%%
\appendix
%%%%%%%%%%%%%%%%%%%%%%%%%%%%%%%%%%%%%%%%%%%%%%%%%%%%%%%%%%

%%%%%%%%%%%%%%%%%%%%%%%%%%%%%%%%%%%%%%%%%%%%%%%%%%%%%%%%%%
\section{\label{app:deriv} Derivation of the scattering matrix}
%\section{Derivation of the scattering matrix}
%%%%%%%%%%%%%%%%%%%%%%%%%%%%%%%%%%%%%%%%%%%%%%%%%%%%%%%%%%

This Appendix derives the scattering matrix introduced in Eq.~$\eqref{eq:scmatrix}$ of Sec.~\ref{sec:model}. We begin from the general single-photon scattering ansatz containing the three independent incoming amplitudes $\alpha_1$, $\alpha_2$, and $\alpha_3$. Substitution into the dynamical equations of Eqs.~$\eqref{eq:dynamicseq}$ and application of the corresponding boundary conditions yields the outgoing amplitudes $\beta_1$, $\beta_2$, and $\beta_3$ as illustrated in Fig.~\ref{fig:fig1-structure}. The resulting expressions are then arranged into the $3\times3$ matrix elements $\bm {S}_{ij}(\varepsilon)$. No additional physical assumptions are introduced in this Appendix.

%%%%%%%%%%%%%%%%%%%%%%%%%%%%%%%%%%%%%%%%%%%%%%%%%%%%%%%%%%
\subsection{General scattering ansatz}
%%%%%%%%%%%%%%%%%%%%%%%%%%%%%%%%%%%%%%%%%%%%%%%%%%%%%%%%%%

To derive the scattering matrix, we use the general single-photon scattering ansatz
\begin{subequations}
\begin{align}
\phi_{R1}(y)&= \alpha_{1}e^{iky}\theta(-y)+\beta_1e^{iky}\theta(y)\\
\phi_{L1}(y)&= \alpha_{2}e^{-iky}\theta(y)+\beta_{2}e^{-iky}\theta(-y)\\
\phi_{R2}(x)&= \beta_{3}e^{ikx}\theta(x)\\
\phi_{L2}(x)&= \alpha_3e^{-ikx}\theta(x)
\end{align}
\label{eq:wavefuncts}
\end{subequations}
where $\theta(x)$ is the unit step function, the waveguide 2 field amplitudes in the scattering ansatz~$\eqref{eq:wavefuncts}$ are
\begin{subequations}
\begin{align}
\phi_{R2}(x)&= \beta_{3}e^{ikx}\theta(x),\\
\phi_{L2}(x)&= \alpha_3e^{-ikx}\theta(x).
\end{align}
\end{subequations}
with the convention $\theta(0)=1/2$, the endpoint fields are $ \phi_{R2}(0)=\beta_3/2$ and $\phi_{L2}(0)=\alpha_3/2$. Equivalently, for a symmetric regularization of the endpoint delta function, $\int_{0}^\infty \delta(x)f(x)dx=f(0)/2$.
%Equivalently, within this prescription, 
%\begin{align}
%\delta(x)\theta(x)=\frac{1}{2}\delta(x).
%\end{align}
These endpoint values explain why the cross-field terms involving $\eta$ acquire factor of 1/2. However, there is an important distinction between evaluating a half-line field at the endpoint and handling the localized source term proportional to $V_2C\delta(x)$.

%%%%%%%%%%%%%%%%%%%%%%%%%%%%%%%%%%%%%%%%%%%%%%%%%%%%%%%%%%
\subsection{Boundary condition at the end-coupled 
waveguide}
%%%%%%%%%%%%%%%%%%%%%%%%%%%%%%%%%%%%%%%%%%%%%%%%%%%%%%%%%%

Waveguide 2 occupies the semi-infinite domain $x\geq 0$ and terminates at $x=0$. The right- and left-propagating field amplitudes must therefore satisfy a boundary condition at this endpoint. To implement this condition within the real-space Hamiltonian, we introduce the auxiliary endpoint interaction proportional to $\eta$ in Eq.~(\ref{eq:componHamil}~c).
\begin{align}
H_{end}=\int_{0}^\infty dx\delta(x)\left[\eta{\hat\psi_{R2}}^\dag (x){\hat\psi_{L2}} (x)+\eta^*{\hat\psi_{L2}}^\dag (x){\hat\psi_{R2}} (x)\right].
\label{eq:hamitonwaveguide2}
\end{align}
The parameter $\eta$ is not an independent cavity-waveguide coupling; its value is fixed by the chosen endpoint reflection condition.
The corresponding endpoint terms must be treated consistently when applying the boundary conditions to the dynamical equations. 
Substitution of the scattering ansatz Eqs.~(\ref{eq:wavefuncts}) into Eq.~(\ref{eq:dynamicseq}~c) and Eq.~(\ref{eq:dynamicseq}~d) gives the endpoint equations
\begin{subequations}
\begin{align}
%&iv_g\alpha_1-iv_g\beta_1+V_1C=0\\
%&iv_g\alpha_2-iv_g\beta_2+V_1C=0\\
&-iv_g\beta_3+\frac{\eta}{2}\alpha_3+V_2C=0\\
&iv_g\alpha_3+\frac{\eta^*}{2}\beta_3+V_2C=0
%&\Omega A+\frac{V_1}{2}[\alpha_1+\beta_1+\alpha_2+\beta_2]+\frac{V_2}{2}[\beta_3+\alpha_3]=(\varepsilon-\tilde{\omega}_c)C\\
%&\Omega C=(\varepsilon-\tilde{\omega}_a)A
\end{align}
\label{eq:endpointequation}
\end{subequations}
In particular, the second equation contains the outgoing amplitude $\beta_3$ in the term proportional to $\eta^*$, because the corresponding interaction in Eq.~(\ref{eq:dynamicseq}~d) couples to the right-propagating field. For the uncoupled-waveguide limit  $V_2=0$, we choose the reflection condition $r_{end}=\beta_3/\alpha_3=+1$. Substitution into Eq.~(\ref{eq:endpointequation}~a) gives 
\begin{align}
\eta=2iv_g
\end{align}
The same value satisfies Eq.~(\ref{eq:endpointequation}~b), confirming consistency of the endpoint condition. 

%%%%%%%%%%%%%%%%%%%%%%%%%%%%%%%%%%%%%%%%%%%%%%%%%%%%%%%%%%
\subsection{Outgoing amplitudes}
%%%%%%%%%%%%%%%%%%%%%%%%%%%%%%%%%%%%%%%%%%%%%%%%%%%%%%%%%%

Substitution Eqs. $\eqref{eq:wavefuncts}$ into Eqs. $\eqref{eq:dynamicseq}$, followed by elimination of the cavity and atomic amplitudes, gives
\begin{widetext}

\begin{subequations}
\begin{align}
&\beta_1=\frac{\varepsilon-\tilde{\omega}_c-\frac{\Omega^2}{\varepsilon-\tilde{\omega}_a}+i\frac{V_2^2}{2v^g}}  {\varepsilon-\tilde{\omega}_c-\frac{\Omega^2}{\varepsilon-\tilde{\omega}_a}+i\frac{V_1^2}{v_g}+i\frac{V_2^2}{2v_g}}\alpha_1
+\frac{\frac{V_1^2}{iv^g}}{\varepsilon-\tilde{\omega}_c-\frac{\Omega^2}{\varepsilon-\tilde{\omega}_a}+i\frac{V_1^2}{v_g}+i\frac{V_2^2}{2v^g}}\alpha_2
+\frac{\frac{V_1V_2}{iv_g}(\frac{\eta}{2iv_g})}{\varepsilon-\tilde{\omega}_c-\frac{\Omega^2}{\varepsilon-\tilde{\omega}_a}+i\frac{V_1^2}{v_g}+i\frac{V_2^2}{2v_g}}\alpha_3\\
&\beta_2=\frac{\frac{V_1^2}{iv_g}}{\varepsilon-\tilde{\omega}_c-\frac{\Omega^2}{\varepsilon-\tilde{\omega}_a}+i\frac{V_1^2}{v_g}+i\frac{V_2^2}{2v_g}}\alpha_1
+\frac{\varepsilon-\tilde{\omega}_c-\frac{\Omega^2}{\varepsilon-\tilde{\omega}_a}+i\frac{V_2^2}{2v_g}}  {\varepsilon-\tilde{\omega}_c-\frac{\Omega^2}{\varepsilon-\tilde{\omega}_a}+i\frac{V_1^2}{v_g}+i\frac{V_2^2}{2v_g}}\alpha_2
+\frac{\frac{V_1V_2}{iv_g}(\frac{\eta}{2iv_g})}{\varepsilon-\tilde{\omega}_c-\frac{\Omega^2}{\varepsilon-\tilde{\omega}_a}+i\frac{V_1^2}{v_g}+i\frac{V_2^2}{2v_g}}\alpha_3\\
&\beta_3=\frac{\frac{V_1V_2}{iv_g}}{\varepsilon-\tilde{\omega}_c-\frac{\Omega^2}{\varepsilon-\tilde{\omega}_a}+i\frac{V_1^2}{v_g}+i\frac{V_2^2}{2v_g}}\alpha_1
+\frac{\frac{V_1V_2}{iv_g}}{\varepsilon-\tilde{\omega}_c-\frac{\Omega^2}{\varepsilon-\tilde{\omega}_a}+i\frac{V_1^2}{v_g}+i\frac{V_2^2}{2v_g}}\alpha_2
+\frac{\frac{\eta}{2iv_g}[\varepsilon-\tilde{\omega}_c-\frac{\Omega^2}{\varepsilon-\tilde{\omega}_a}+i\frac{V_1^2}{v_g}]+\frac{V_2^2}{2iv_g} }{\varepsilon-\tilde{\omega}_c-\frac{\Omega^2}{\varepsilon-\tilde{\omega}_a}+i\frac{V_1^2}{v_g}+i\frac{V_2^2}{2v_g}}\alpha_3
\end{align}
\label{eq:scatsol2}
\end{subequations}

\end{widetext}

%%%%%%%%%%%%%%%%%%%%%%%%%%%%%%%%%%%%%%%%%%%%%%%%%%%%%%%%%%
\subsection{Effective coupling parameters}
%%%%%%%%%%%%%%%%%%%%%%%%%%%%%%%%%%%%%%%%%%%%%%%%%%%%%%%%%%

Introducing the effective coupling parameters $J_1=V_1^2/v_g$ and $J_2=V_2^2/v_g$, the outgoing amplitudes can be written in terms of the common denominator $D(\varepsilon)$ as  
\begin{subequations}
\begin{align}
\beta_1=&\frac{A(\varepsilon)+i\frac{J_2}{2}}  {D(\varepsilon)}\alpha_1
+\frac{-iJ_1}{D(\varepsilon)}\alpha_2
+\frac{-i\sqrt{J_1J_2}(\frac{\eta}{2iv_g})}{D(\varepsilon)}\alpha_3\\
\beta_2=&\frac{-iJ_1}{D(\varepsilon)}\alpha_1
+\frac{A(\varepsilon)+i\frac{J_2}{2}}  {D(\varepsilon)}\alpha_2
+\frac{-i\sqrt{J_1J_2}(\frac{\eta}{2iv_g})}{D(\varepsilon)}\alpha_3\\
\beta_3=&\frac{-i\sqrt{J_1J_2}}{D(\varepsilon)}\alpha_1
+\frac{-i\sqrt{J_1J_2}}{D(\varepsilon)}\alpha_2
+\frac{\frac{\eta}{2iv_g}[A(\varepsilon)+iJ_1]-i\frac{J_2}{2} }{D(\varepsilon)}\alpha_3
\end{align}
\label{eq:scatsol3}
\end{subequations}
%\begin{widetext}
%\begin{subequations}
%\begin{align}
%&\beta_1=\frac{\varepsilon-\tilde{\omega}_c-\frac{\Omega^2}{\varepsilon-\tilde{\omega}_a}+i\frac{J_2}{2}}  {D(\varepsilon)}\alpha_1
%+\frac{-iJ_1}{D(\varepsilon)}\alpha_2
%+\frac{-i\sqrt{J_1J_2}\left(\frac{\frac{\eta}{2iv_g}+1}{2}\right)}{D(\varepsilon)}\alpha_3\\
%&\beta_2=\frac{-iJ_1}{D(\varepsilon)}\alpha_1
%+\frac{\varepsilon-\tilde{\omega}_c-\frac{\Omega^2}{\varepsilon-\tilde{\omega}_a}+i\frac{J_2}{2}}  {D(\varepsilon)}\alpha_2
%+\frac{-i\sqrt{J_1J_2}\left(\frac{\frac{\eta}{2iv_g}+1}{2}\right)}{D(\varepsilon)}\alpha_3\\
%&\beta_3=\frac{-i\sqrt{J_1J_2}}{D(\varepsilon)}\alpha_1
%+\frac{-i\sqrt{J_1J_2}}{D(\varepsilon)}\alpha_2
%+\frac{\frac{\eta}{2iv_g}[\varepsilon-\tilde{\omega}_c-\frac{\Omega^2}{\varepsilon-\tilde{\omega}_a}+iJ_1]-i\frac{J_2}{2} }{D(\varepsilon)}\alpha_3
%\end{align}
%\label{eq:scatsol3}
%\end{subequations}
%\end{widetext}
where
\begin{align}
D(\varepsilon)=A(\varepsilon)+iJ_1+\frac{iJ_2}{2}
\label{eq:asympoleparameter}
\end{align}
is the pole parameter that incorporates the cavity self-energy.
%\begin{align}
%D(\varepsilon)=\varepsilon-\tilde{\omega}_c+iJ_1+\frac{iJ_2}{2}-\frac{\Omega^2}{\varepsilon-\tilde{\omega}_a}
%\label{eq:asympoleparameter}
%\end{align}

%%%%%%%%%%%%%%%%%%%%%%%%%%%%%%%%%%%%%%%%%%%%%%%%%%%%%%%%%%
\subsection{Scattering matrix}
%%%%%%%%%%%%%%%%%%%%%%%%%%%%%%%%%%%%%%%%%%%%%%%%%%%%%%%%%%

Collecting the outgoing amplitudes into the vector $\bm \beta_{out}$ gives the scattering relation
\begin{align}
\bm \beta_{out}=\bm {S}(\varepsilon)\bm \alpha_{in} 
\label{eq:scatterma}
\end{align}
where the matrix elements are 
\begin{align}
\bm {S}(\varepsilon)
=\frac{1}{D(\varepsilon)}\left(\begin{array}{ccc}
  A(\varepsilon)+\frac{iJ_2}{2} &  -iJ_1 &  -i\sqrt{J_1J_2} \\
  -iJ_1 & A(\varepsilon)+\frac{iJ_2}{2} &  -i\sqrt{J_1J_2} \\
  -i\sqrt{J_1J_2} &  -i\sqrt{J_1J_2} &  A(\varepsilon)+iJ_1-\frac{iJ_2}{2}
\end{array}\right)
\label{eq:scatteringmatrix2}
\end{align}
with 
\begin{align}
A(\varepsilon)=\varepsilon-\tilde{\omega}_c-\frac{\Omega^2}{\varepsilon-\tilde{\omega}_a}
\label{eq:aparameter}
\end{align}
%the matrix structure consistent with coupling definitions as
%\begin{subequations}
%\begin{align}
 %S_{11}&= S_{22} = \frac{\varepsilon-\tilde{\omega}_c-\frac{\Omega^2}{\varepsilon-\tilde{\omega}_a}+i\frac{J_2}{2}}  {D(\varepsilon)}\\
 %S_{12}&= S_{21} = -\frac{iJ_1}{D(\varepsilon)}\\
 %S_{13}&=  S_{23} = -\frac{i\sqrt{J_1J_2}\frac{\frac{\eta}{2iv_g}+1}{2}}{D(\varepsilon)}\\
 %S_{31}&=  S_{32} = -\frac{i\sqrt{J_1J_2}}{D(\varepsilon)}\\
 %S_{33}&= \frac{\frac{\eta}{2iv_g}[\varepsilon-\tilde{\omega}_c-\frac{\Omega^2}{\varepsilon-\tilde{\omega}_a}+iJ_1]-i\frac{J_2}{2} }{D(\varepsilon)}
%\end{align}
%\label{eq:scoefici}
%\end{subequations}
%where $J_1$ and $J_2$ are the two effective cavity-waveguide coupling strengths, which may be unequal in the asymmetric configuration. 
At the minimum-coupling case $J_1=J_2=0$, one can find $\bm {S}(\varepsilon)$  Eq. $\eqref{eq:scatteringmatrix2}$ as,
\begin{align}
\bm {S}(\varepsilon)=\left(\begin{array}{ccc}
 1&0&0 \\
 0&1&0 \\
 0&0&\frac{\eta}{2iv^g}
\end{array}\right)
\end{align}
chosen the physical endpoint reflection condition $r_{end}=+1$, gives $\eta=2iv_g$. Then the resulting lossless $\bm {S}(\varepsilon)$ matrix is unitary.

%%%%%%%%%%%%%%%%%%%%%%%%%%%%%%%%%%%%%%%%%%%%%%%%%%%%%%%%%%
\subsection{Symmetric-coupling limit}
%%%%%%%%%%%%%%%%%%%%%%%%%%%%%%%%%%%%%%%%%%%%%%%%%%%%%%%%%%

Now, we use symmetric case $J_1=J_2=J$ explicitly to check and verify asymmetric effective coupling strengths in the scattering matrix approach. With these expressions, the symmetric limit becomes especially simple
\begin{align}
\bm {S}(\varepsilon)
=\frac{1}{D(\varepsilon)}\left(\begin{array}{ccc}
  A(\varepsilon)+\frac{iJ}{2} &  -iJ &  -iJ \\
  -iJ & A(\varepsilon)+\frac{iJ}{2} &  -iJ \\
  -iJ &  -iJ &  A(\varepsilon)+\frac{iJ}{2}
\end{array}\right)
\label{eq:scatteringmatrixsymm}
\end{align}
with
\begin{align}
D(\varepsilon)=A(\varepsilon)+\frac{3iJ}{2}
\label{eq:sympoleparameter}
\end{align}

%%%%%%%%%%%%%%%%%%%%%%%%%%%%%%%%%%%%%%%%%%%%%%%%%%%%%%%%%%
\subsection{Scattering probabilities}
%%%%%%%%%%%%%%%%%%%%%%%%%%%%%%%%%%%%%%%%%%%%%%%%%%%%%%%%%%

Finally, by applying the general formula for scattering probabilities Eq. $\eqref{eq:genproba}$, we can obtain
\begin{subequations}
\begin{align}
P_1=|\beta_1|^2=|S_{11}\alpha_1+S_{12}\alpha_2+S_{13}\alpha_3|^2\\
P_2=|\beta_2|^2=|S_{21}\alpha_1+S_{22}\alpha_2+S_{23}\alpha_3|^2\\
P_3=|\beta_3|^2=|S_{31}\alpha_1+S_{32}\alpha_2+S_{33}\alpha_3|^2
\end{align}
\end{subequations}
the elements of matrix $\bm {S}_{ij}(\varepsilon)$ in the case of asymmetry $J_1\neq J2$ are

\begin{subequations}
\begin{align}
 S_{11}&=  S_{22} = \frac{ A(\varepsilon)+i\frac{J_2}{2}}  {D(\varepsilon)}\\
 S_{12}&=  S_{21} = -\frac{iJ_1}{D(\varepsilon)}\\
 S_{13}&=  S_{23} = S_{31}=  S_{32}=-\frac{i\sqrt{J_1J_2}}{D(\varepsilon)}\\
 %S_{31}&=  S_{32} = -\frac{i\sqrt{J_1J_2}}{D(\varepsilon)}\\
 S_{33}&= \frac{ A(\varepsilon)+iJ_1-i\frac{J_2}{2} }{D(\varepsilon)}
\end{align}
\label{eq:scoeficij1notj2}
\end{subequations}
with
\begin{align}
D(\varepsilon)=A(\varepsilon)+iJ_1+\frac{iJ_2}{2}
%\label{eq:asympoleparameter}
\end{align}
whereas in a symmetric case are 
\begin{subequations}
\begin{align}
 S_{ij}&= \frac{ A(\varepsilon)+iJ/2}{D(\varepsilon)}, (i=j),\\
 S_{ij}&= -\frac{iJ}{D(\varepsilon)}, (i\neq j).
\end{align}
\label{eq:scoeficij}
\end{subequations}
with
\begin{align}
D(\varepsilon)=A(\varepsilon)+\frac{3iJ}{2}
%\label{eq:sympoleparameter}
\end{align}
\bibliography{references} 

\end{document}